\documentclass[journal]{IEEEtran}

\ifCLASSINFOpdf
\else
\fi

\usepackage{amsmath}
\usepackage[pdftex]{graphicx}
\usepackage{amssymb}
\usepackage{color}

\usepackage{algorithm}
\usepackage{algorithmic}
\usepackage{enumerate}
\usepackage{subcaption}
\usepackage{enumitem}
\usepackage{cite}

\DeclareMathOperator\supp{supp}
\DeclareMathOperator*{\argmin}{argmin}

\usepackage{amsthm}% http://ctan.org/pkg/amsthm         
\makeatletter
\newtheoremstyle{break}
   {\topsep}{\topsep}%
   {\itshape}{}%
   {\bfseries}{}%
   {\newline}
   {\thmname{#1}\thmnumber{\@ifnotempty{#1}{ }\@upn{#2}}%
    \thmnote{ {\bfseries(#3)}}}% 
\makeatother
\theoremstyle{break}

\newtheorem{theorem}{Theorem}%[section]
\newtheorem{lemma}[theorem]{Lemma}

\newtheorem{corollary}[theorem]{Corollary}
\newtheorem{definition}{Definition}

\begin{document}

\title{Blue-Noise Sampling on Graphs}

\author{Alejandro~Parada-Mayorga,
        Daniel L.~Lau,~\IEEEmembership{Senior Member,~IEEE},
        Jhony H. Giraldo, and
        Gonzalo R.~Arce,~\IEEEmembership{Fellow,~IEEE,} 
        % <-this % stops a space
\thanks{Alejandro Parada-Mayorga and Jhony H. Giraldo are with the Department
of Electrical and Computer Engineering, University of Delaware, Newark,
DE, 19716 USA e-mail: alejopm@udel.edu, jgiraldo@udel.edu.}
\thanks{Daniel L. Lau is with  the Department
of Electrical and Computer Engineering, University of Kentucky, Lexington, Kentucky, 40526 USA e-mail: dllau@uky.edu.}
\thanks{Gonzalo R. Arce is with Institute for Financial Services Analytics and the Department
	of Electrical and Computer Engineering, University of Delaware, Newark,
	DE, 19716 USA e-mail:arce@udel.edu.}
\thanks{This work was supported in part by the National Science Foundation, grant NSF CFI1815992, by the UDRF foundation under Strategic Initiative Award, and by a University of Delaware Dissertation Fellowship Award.}
}

\markboth{Signal Processing}%
{Shell \MakeLowercase{\textit{et. al.}}: Bare Demo of IEEEtran.cls for Journals}

\maketitle

\begin{abstract}
In the area of graph signal processing, a graph is a set of nodes arbitrarily connected by weighted links; a graph signal is a set of scalar values associated with each node; and sampling is the problem of selecting an optimal subset of nodes from which 
a graph signal can be reconstructed. This paper proposes the use of spatial dithering on the vertex domain of the graph, as a way to conveniently find statistically good sampling sets. This is done establishing that there is a family of good sampling sets
characterized on the vertex domain by a maximization of the distance between sampling nodes; in the Fourier domain, these are characterized by spectrums that are dominated by high frequencies referred to as blue-noise. The theoretical connection
between blue-noise sampling on graphs and previous results in graph signal processing is also established, explaining the advantages of
the proposed approach. Restricting our analysis to undirected and connected graphs, numerical tests are performed in order to compare the effectiveness of blue-noise sampling against other approaches.
\end{abstract}

% Note that keywords are not normally used for peer review papers.

\IEEEpeerreviewmaketitle

\section{Introduction}
Interesting phenomena in nature can often be captured by graphs since objects and data are invariably inter-related in some sense. Social~\cite{roberts1978graph}, financial~\cite{KELLER2007469}, ecological networks, and the human
brain~\cite{fornito2016fundamentals} are a few examples of such networks. Data in these networks reside on irregular or otherwise unordered structures~\cite{barabasi2016network}. New data science tools are thus emerging to process signals on graph
structures where concepts of algebraic and spectral graph theory are being merged with methods used in computational harmonic analysis~\cite{shuman_magazine,shumanvertexfrequanly,shuman2016multiscale}. A common problem in these networks is to 
determine which nodes play the most important role, assuming there is a quantity of interest defined on the network. 

Graph signal sampling thus becomes essential. Naturally, the mathematics of sampling theory and spectral graph theory have been combined leading to generalized Nyquist sampling
principles for graphs~\cite{shuman_magazine,pesensonschrodinger,pesenson2010,ortega_gsp,6638704,7581102}.  In general, these methods are based on the underlying graph spectral decompositions~\cite{7581102,anis_conf,chensamplingongraphs,bookgspchapsampling}.

This work explores a somewhat radical departure from prior work, inspired by sampling patterns in 
traditional dithering and halftoning.
Specifically, 
we intend to design graph signal sampling techniques that promote the maximization of the distance between sampling nodes on the vertex domain that are typically characterized by a low frequency energy. Sampling patterns with these characteristics 
are referred to in the spatial dithering literature as~\textit{blue-noise}~\cite{ulichney,bluenoiseLaumagazine}.

In this paper, the connection between the properties of blue-noise sampling patterns and the results related with sampling sets in graphs is established, showing that blue-noise like sampling patterns in graphs are connected with good sampling sets 
in terms of preserving the uniqueness of the representation of the sampled signal in a noise-free scenario. Additionally, it is shown how the inter-distance between the sampling nodes affects 
the redness in a given sampling pattern. We provide a measure of the bandwidth of the signals that can be uniquely represented from the vertex-domain characteristics of blue-noise sampling patterns. 
A numerical algorithm is proposed in order to compute these blue-noise patterns based on their vertex-domain distribution. In 
particular, trying to exploit the  distribution of the sampling points on the nodes of the graph, a void and cluster algorithm on graphs is developed~\cite{ulichney_voidcluster}, allowing the generation of patterns that lead to reconstruction errors
of bandlimited signals, similar to the ones obtained in the state-of-the-art literature.

This work is organized as follows: in Section II, notation and basic concepts about signal processing on graphs are stated presenting also a description of previous approaches about sampling on graphs. In  Section III blue-noise sampling on graphs is
discussed. In Section IV an algorithm for the generation of blue-noise patterns is discussed, whereas in Section V a set of numerical tests show the performance of the proposed algorithm against other techniques. Finally, in Section  VI a set of conclusions 
is presented.

\section{Preliminaries}
Sandryhaila~\cite{dsp_graphsmoura} proposed a theoretical framework for the analysis and processing of signals on graphs based on the properties of the adjacency matrix. This approach is rooted in \textit{algebraic signal processing}, whereas authors
like Fuhr and Pesenson~\cite{pesensonschrodinger, pesenson2010,fuhrpesenson}, Puy~\cite{puysampling} and Shuman~\cite{shuman_magazine, shumanvertexfrequanly, shumanmultscalepyr}  based their analysis of signals on graphs, relying on the properties 
of the \textit{Laplacian matrix}. In both approaches the Fourier transform of the signals on the graph is defined in terms of a spectral decomposition of the adjacency matrix and the Laplacian matrix respectively, using the set 
of eigenvectors as the Fourier basis for the representation of the signals. 

The first approach offers a direct connection with the shift operator used in traditional signal processing, while the second resembles the main ideas of Fourier analysis in linear spaces in which the eigenfunctions of the Laplacian operator are used 
as the basis representation of the signal. The two approaches use a unitary operator, and problems like sampling and filtering can be successfully considered in both scenarios. In this work,  the combinatorial Laplacian matrix is used as the building 
block, and the graphs considered are undirected, weighted, connected and simple. Consequently, part of the developments proposed rely on the theoretical results obtained by Furh and Pesenson~\cite{pesensonschrodinger, pesenson2010,fuhrpesenson} in 
harmonic analysis on graphs.

\subsection{Graph Signal Sampling}
Let $G=\left(V(G),E(G)\right)$ be an undirected, weighted, connected, simple graph with a set of nodes, $V(G)$, and a set of edges, $E(G)$. $\mathbf{W}$ is the adjacency matrix (symmetric), with $\mathbf{W}(u,v)\geq 0$ the weight connecting the nodes $u$ and $v$ and $u\sim v$ indicates that $\mathbf{W}(u,v)>0$. The degree matrix, $\mathbf{D}$, is a diagonal matrix whose entries are given according to:
\begin{equation}
\mathbf{D}(u,u)=\sum_{v\in V(G)}\mathbf{W}(u,v). 
\end{equation}
For any graph $G$, its volume is defined as $vol(G)=\sum_{u\in V(G)}\mathbf{D}(u,u)$, and the volume of a subset $S\subset V(G)$  is defined as $vol(S)=\sum_{u\in S}\mathbf{D}(u,u)$. On the graph $G$, the combinatorial Laplacian operator is defined as the positive semi-definite operator:
\begin{equation}
\mathbf{L}=\mathbf{D}-\mathbf{W}, 
\end{equation}
whose eigenvalues are organized as $0\leq\mu_{1}\leq\mu_{2}\leq\ldots\leq\mu_{N}$, $N=\vert V(G)\vert$~\cite{biyikoglu2007laplacian}.  
A real signal, $\boldsymbol{x}$, on the graph is then defined as the mapping $\boldsymbol{x}:V(G)\longrightarrow\mathbb{R}$ 
represented by the vector $\boldsymbol{x}\in\mathbb{R}^{N}$ where $\boldsymbol{x}(v)$ is the value of the signal associated 
to $v\in V(G)$. The support of $\boldsymbol{x}$ is represented as $\supp(\boldsymbol{x})$, and the restriction of 
$\boldsymbol{x}$, to any subset $S\subset V(G)$, is represented as $\boldsymbol{x}(S)$. It is worth noticing that:
\begin{equation}
(\mathbf{L}\boldsymbol{x})(v)=\sum_{u\in V(G)}\left( \boldsymbol{x}(v)-\boldsymbol{x}(u)\right)\mathbf{W}(v,u). 
\end{equation}

If the spectral decomposition of the operator $\mathbf{L}$ is represented as $\mathbf{L}=\mathbf{U}\boldsymbol{\Lambda}\mathbf{U}^{\mathsf{T}}$,  then the Graph Fourier Transform~(GFT) of the signal, $\boldsymbol{x}$ on $G$, is given
by $\hat{\boldsymbol{x}}=\mathbf{U}^{\mathsf{T}}\boldsymbol{x}$. There is a direct analogy between the concept of frequency in traditional Fourier Analysis and the behavior of the Graph Fourier Transform as is stated in~\cite{shuman_magazine}. Considering 
this analogy, the bandwidth of a signal $\boldsymbol{x}$ can be defined using the nonzero components of $\hat{\boldsymbol{x}}$.
It is said that $\boldsymbol{x}$ has bandwidth
$\omega\in\mathbb{R}_{+}$ on the spectral axes if $\hat{\boldsymbol{x}}\in PW_{\omega}(G)=\text{span}\{ \mathbf{U}_{k}: \mu_{k}\leq\omega\}$, where $PW_{\omega}(G)$ is the Paley-Wiener space of bandwidth $\omega$~\cite{pesensonschrodinger} 
and $\mathbf{U}_{k}$ indicates the first $k$ column vectors in $\mathbf{U}$. In some cases the bandwidth is also represented with the largest integer $k$ such that $\mu_{k}\leq\omega$.

Given a notion of bandwidth, one invariably questions the notion of sampling rate and whether the number of samples or nodes of a graph can be reduced without loss of information to the signal. We, therefore, define sampling of a
signal $\boldsymbol{x}$ on the graph $G$, by choosing the components of $\boldsymbol{x}$ on a subset of nodes, $S=\{s_{1},\ldots,s_{m} \}\subset V(G)$.  The sampled signal is given by
$\boldsymbol{x}(S)=\mathbf{M}\boldsymbol{x}$ where $\mathbf{M}$ is a binary matrix whose entries are given by $\mathbf{M}=[\boldsymbol{\delta}_{s_{1}},\ldots,\boldsymbol{\delta}_{s_{m}}]^{\mathsf{T}}$ and
$\boldsymbol{\delta}_{v}$ is the $N-$ dimensional Kronecker column vector centered at $v$. Given  $\boldsymbol{x}(S)$, it is possible to obtain a reconstructed version of $\boldsymbol{x}$ in different ways depending on
whether the bandwidth of  the signal is known. We assume that the bandwidth is known  and that  the reconstruction is given by:
\begin{equation}
\boldsymbol{x}_{rec}=\argmin_{\boldsymbol{z}\in span(\mathbf{U}_{k})} \Vert \mathbf{M}\boldsymbol{z}-\boldsymbol{x}(S)\Vert^{2}_{2}=\mathbf{U}_{k}\left( \mathbf{M}\mathbf{U}_{k}\right)^{\dagger}\boldsymbol{x}(S) 
\label{eq_basicrec}
\end{equation}
where $\left( \mathbf{M}\mathbf{U}_{k}\right)^{\dagger}$ is the Moore-Penrose pseudo-inverse of  $\mathbf{M}\mathbf{U}_{k}$~\cite{ortega_proxies,tremblayAB17}.  Alternatively,  in~\cite{pesenson2009} it is shown that a consistent reconstruction of the signal can be obtained from its samples using \textit{interpolation splines}.

The problem of optimally sampling a signal on a graph can now be summarized as choosing $S$ such that we maximize the available bandwidth of $\boldsymbol{x}(S)$. To this end, Pesenson defines~\cite{pesensonschrodinger,pesenson2010} 
a $\Lambda$-removable set for $\Lambda>0$ as the subset of nodes, $S\subset V(G)$, for which:
\begin{equation}
\Vert \boldsymbol{x}\Vert_{2} \leq (1/\Lambda)\Vert \mathbf{L}\boldsymbol{x}\Vert_{2} \quad\quad\forall~\boldsymbol{x}\in L_{2}(S),
\label{eq_Lambdaremsetdef}
\end{equation}
where $L_{2}(S)$ is the set of all signals, $\boldsymbol{x}$, with support in $S\subset V(G)$ (i.e. elements of $\boldsymbol{x}$ not included in $S$ are equal to zero) and
finite $\ell_{2}$ norm. The largest value of $\Lambda$ for which eqn.~(\ref{eq_Lambdaremsetdef}) holds is denoted by $\Lambda_{S}$.
Notice that for any subset of nodes there exists a $\Lambda$-removable set with larger or smaller $\Lambda_{S}$. Therefore, $\Lambda_{S}$ ultimately determines how much importance a given set has in the sampling process of a signal with a specific bandwidth. The relationship between properties of removable sets and the sampling problem was established by Pesenson in the following 
theorem:
\begin{theorem}[Theorem~5.1 in~\cite{pesensonschrodinger}]
If for a set $S\subset V(G)$, its compliment $S^{c}=V(G)\setminus S$ is a $\Lambda_{S^{c}}-$removable set, then all signals in $PW_{\omega}(G)$ are completely determined by its values in $S$, whenever \linebreak $0<\omega<\Lambda_{S^{c}}$.
\label{theorem_pensensonuniquesets}
\end{theorem}
\noindent In~\cite{fuhrpesenson}, another result related with sampling sets is established using a constant that can be calculated directly with the weights, $\mathbf{W}$, of the graph, $G$, stated in the following theorem:
\begin{theorem}[\cite{fuhrpesenson}]
Every $S\subset V(G)$ is a uniqueness set for all functions in $PW_{\omega}(G)$ with any $\omega<K_{S}$, where 
\begin{equation}
K_{S}=\inf_{v\in S^{c}}w_{S}(v)
\end{equation}
and $w_{S}(v)=\sum_{s\in S}\mathbf{W}(s,v)$.
\label{theorem_Kstheory}
\end{theorem}
\noindent Theorems~\ref{theorem_pensensonuniquesets} and~\ref{theorem_Kstheory} play a central role in the description of properties for different classes of sampling sets as it is possible to consider that a \textit{good} sampling  set, $S$, promotes the 
maximization of constants, $\Lambda_{S^{c}}$ and $K_{S}$. In particular, it will be shown in the following sections that blue-noise sampling patterns indeed promote high values of these constants.

Recently Pesenson~\cite{pesenson2019} introduced results that characterize the representation of a band limited signal in terms of induced subgraphs obtained from partitions of $V(G)$ that cover $V(G)$. 
% More specifically, 
% let $\mathcal{P}=\left\lbrace V(\Omega_{1}), V(\Omega_{2}), \ldots, V(\Omega_{\vert\mathcal{P}\vert})\right\rbrace$ be a disjoint cover of $V(G)$ by \textit{connected subgraphs} $\Omega_{j}$, and lets consider the functions
% \begin{equation}
% \xi_{j}=\frac{1}{\sqrt{\vert V(\Omega_{j})\vert}}\chi_{j},\quad  
% \zeta_{j}=\frac{1}{\vert V(\Omega_{j})\vert}\chi_{j}
% \end{equation}
% with $\chi_{j}$ being the characteristic function of $V(\Omega_{j})$. Let $\mu_{1,j}$ be the first nonzero eigenvalue of $\mathbf{L}_{\Omega_{j}}$, where $\mathbf{L}_{\Omega_{j}}$ is the Laplacian on the \textit{induced subgraph} $\Omega_{j}$. Then, it is shown in 
% ~\cite{pesenson2019} that if $\Lambda_{\mathcal{P}}=\inf_{j}\mu_{1,j}>0$, then the set of functions $\{ \zeta_{j}\}_{j=1}^{\vert\mathcal{P}\vert}$ is a frame for any space $PW_{\omega}(G)$ as long as $\Lambda_{\mathcal{P}}>\frac{1+\alpha}{\alpha}\omega$, with
% $\alpha>0$. With these conditions, Pesenson showed that any signal in $PW_{\omega}(G)$ can be uniquely represented and determined from its average values on the elements of the partition $\mathcal{P}$. 
This statement can be summarized in the following theorem. 

\begin{theorem}[5.1,6.1,6.2 \cite{pesenson2019}]
Let $G$ be a connected finite or infinite and countable graph. Suppose that $\mathcal{P}=\{V(\Omega_{j})\}_{j=1}^{j=\vert \mathcal{P}\vert}$ is a disjoint cover of $V(G)$ by \textbf{connected} and finite subgraphs $\Omega_{j}$. Let
$\mathbf{L}_{\Omega_{j}}$ be the Laplace operator of the \textbf{induced} graph $\Omega_{j}$ whose first nonzero eigenvalue is $\mu_{1,j}$. If $\Lambda_{\mathcal{P}}=\inf_{j}\mu_{1,j}>0$ and $\Lambda_{\mathcal{P}}>\frac{1+\alpha}{\alpha}\omega$ with $\alpha>0$, then
every signal $\boldsymbol{x}\in PW_{\omega}(G)$ is uniquely determined by the values
$\boldsymbol{x}^{\mathsf{T}}\boldsymbol{\xi}_{j}$, where $\boldsymbol{\xi}_{j}=\boldsymbol{\chi}_{j}/\sqrt{\vert V(\Omega_{j})\vert}$ with $\boldsymbol{\chi}_{j}(V(\Omega_{j}))=1$  and $\boldsymbol{\chi}_{j}(V(\Omega_{j})^{c})=0$.  Additionally, $\boldsymbol{x}$ can be reconstructed from this set of values in a stable way.
\label{pesentheorem2019_3}
\end{theorem}
\noindent It is important to remark the meaning and implications of Theorem~\ref{pesentheorem2019_3}. This result shows that $V(G)$ can be divided into disjoint subsets that 
cover $V(G)$, and a given band limited signal can be reconstructed
from the average values of the signal in those regions. Additionally, the constant $\Lambda_{\mathcal{P}}$ associated to the partition provides  a measure of the quality of the reconstruction obtained from the regions on $V(G)$ defined by $\mathcal{P}$.
It is also worthy to point out that the size of the elements in the partition has a natural limit as $\mathbf{L}_{\Omega_{j}}$ is expected to have at least one nonzero eigenvalue, which would not be the case
when $\Omega_{j}$ consist of one single vertex. This result will allow us to establish a connection between the spectral and vertex domain behavior of sampling patterns in some classes of graphs. Additionally, we will show that from a blue-noise sampling
pattern
$\boldsymbol{s}$, 
it is possible to build a partition that can be used to estimate the bandwidth of signals that are uniquely represented  by their samples on the sampling nodes indicated by $\boldsymbol{s}$.

\subsection{Optimal Graph Sampling}
The problem of finding the best $S$ is a combinatorial problem of calculating $\Lambda_{S^c}$ for all sampling sets and choosing the set with the largest value of $\Lambda_{S^{c}}$, a prohibitively expensive  process 
for large graphs. Allowing for some short cuts, a simple, greedy procedure for finding a good sampling set starts with an empty set of nodes and iteratively adds one node at a time, taking the best available node at each iteration according the value of a 
cost function. Several authors have formulated the problem of sampling and reconstruction in the presence of measurement noise, and in these works objective functions have been proposed that minimize the reconstruction error in terms 
of the worst case~\cite{chensamplingongraphs}, where
$S^{opt}=\arg\max_{\vert S\vert=m}\sigma_{1}^{2}$,
the mean case~\cite{ortega_proxies}, 
where $S^{opt}=\arg\max_{\vert S\vert=m}\sum_{i=1}^{\min \{m,k\}} \sigma_{i}^{-2}$, 
and the maximum volume case~\cite{tremblayAB17}, 
where $S^{opt}=\arg\max_{\vert S\vert=m}\prod_{i=1}^{\min \{m,k\}}\sigma_{i}^{2}$; and $\sigma_{i}$ represents the $i^{th}$ singular value of the matrix $\mathbf{M}\mathbf{U}_{k}$ consisting of the first $k$ eigenvectors of ${\bf W}$ or ${\bf L}$ respectively,
sampled on the rows indicated by $S$. In~\cite{tsitsverobarbarossa} the optimal sampling set is obtained considering the same cost function for the mean case, but using the singular values 
of $\mathbf{\Lambda}\mathbf{U}^{\mathsf{T}}\text{diag}(S)$, where
$\text{diag}(S)$ is the diagonal matrix whose entries are given by $\text{diag}(S)_{i,i}=1\Leftrightarrow i\in S$.

In order to reduce computational complexity, Anis et. al.~\cite{ortega_proxies} defines {\it graph spectral proxies} of order $q$ as estimates of the cutoff frequency of a given signal which can be used to define cutoff frequency estimates for a subset 
of nodes $S$ according to:
%%%%%%%%%%%%%%%%%%
\begin{equation}
\Omega_q(S) =\min_{\phi\in L_{2}(S^{c})}\left( \frac{\Vert{\bf L}^{k}\phi\Vert_{2}}{\Vert \phi\Vert_{2}} \right)^{\frac{1}{q}},
\label{cutoffEstimateEqn}
\end{equation}
%%%%%%%%%%%%%%%%%%
with $\mathbf{L}^{q}$ being the $q^{th}$ power of $\mathbf{L}$~\cite{ortega_proxies,anis_conf}. Anis et. al. further shows that, for any $q\in\mathbb{N}$ and $S$, it is possible to have perfect reconstruction when $\omega<\Omega_q(S)$. The 
value of  $\Omega_q(S)$ can be calculated as $\Omega_q(S)=( \sigma_{1,q})^{\frac{1}{2q}}$, where $\sigma_{1,q}$ denotes the smallest eigenvalue of the reduced matrix 
$\mathbf{L}^{2q}_{S^{c},S^{c}}$. The optimal sampling 
set can then be represented as the solution of the problem:
\begin{equation}
S_{q}^{opt}=\arg\max_{\vert S\vert=m} \Omega_q(S),
\label{eq_combsearch2ortega}
\end{equation}
which is still combinatorial; however, Anis et. al. proposes a heuristic rule to solve eqn.~(\ref{eq_combsearch2ortega}) using the first eigenvector of ${\bf L}^{q}_{S^{c},S^{c}}$. Basically, a 
node is added 
to the sampling set according to the index of the component with maximum absolute value for the first eigenvector of ${\bf L}^{q}_{S^{c},S^{c}}$. The quality of the sampling set is also related to the value of $q$, which should be 
selected as large as possible at the expense of a higher computational cost. In~\cite{8047995}, some performance theoretical bounds for these greedy sampling techniques are derived.

In some scenarios for sampling signals on graphs a spectral decomposition of the operators is not available, and therefore, there is a strong need for vertex domain sampling schemes that attempt to build good sampling patterns based entirely on the local graph
structure around a node. In particular, for those cases where the graphs are too large for calculating the eigenvalues and eigenvectors of the GFT, several authors have looked at the problem of sampling using subsets of nodes that may not be optimal but are still very good at preserving band-limited signals. In the case of Puy et. al~\cite{puysampling}, the authors perform a random selection of nodes with a recovery algorithm that involves a  probability distribution on a diagonal matrix, ${\bf P}$, in addition to the sampling
matrix operator $\mathbf{M}$. The reconstructed signal $\boldsymbol{x}_{rec}$ can then be calculated as:
\begin{equation}
\boldsymbol{x}_{rec}=\arg\min_{\boldsymbol{z}\in\mathbb{R}^{N}}\left(\left\Vert {\bf P}^{-1/2}({\bf M}\boldsymbol{z}-\boldsymbol{x}(S))\right\Vert^{2}_{2}+\tau\boldsymbol{z}^\intercal{g(\bf L})\boldsymbol{z}\right),
\label{eq_puyreconst}
\end{equation}
where $\boldsymbol{x}(S)={\bf M}\boldsymbol{x}$ is the sampled version of the signal $\boldsymbol{x}$, $\tau$ is a regularization parameter selected empirically and $g(\cdot)$ is a polynomial function selected also empirically. 

Puy~{\it et. al}~\cite{puysampling} further show that an optimal ${\bf P}$ can be determined by the use of the \textit{local graph coherence}, $\nu_{k}$, on the nodes of the graph. The value of $\nu_{k}(i)$ at the node $i$ can be calculated as $\nu_{k}(i)=\Vert {\bf U}_{k}\boldsymbol{\delta}_{i}\Vert_{2}$, where $\boldsymbol{\delta}_{i}$ is the Kronecker vector centered at node $i$, and it provides a measure about how important is the node $i$ for the sampling of a signal with bandwidth $k$. If $\nu_{k}(i)$ is equal to 1 for a particular node $i$, then there exists $k$-bandlimited graph signals whose energy is solely concentrated in this $i^{th}$ node.  If $\nu_{k}(i)$ is equal to 0, then no $k$-bandlimited graph signal has any energy in this $i^{th}$ node. Therefore, node $i$ can be deleted with no repercussions. 

Because the calculation of $\nu_{k}(i)$ requires the knowledge of the spectral decomposition, Puy~{\it et. al} propose an approximate estimation of $\nu_{k}(i)$ that can be obtained without the calculation of any spectral decomposition, which allows the solution of eqn.~(\ref{eq_puyreconst}). When the optimal ${\bf P}$ is used, Puy~{\it et. al} show that the matrix ${\bf M}{\bf P}^{-1/2}$ satisfies a restricted isometry property when the number of samples is on the order of $O(k\log k)$, which provides a strong guarantee for the exact recovery of the signal. This represents an elegant result but with the drawback that $O(k\log k)$ is substantially higher than $k$, which is the optimal number of samples required to reconstruct a signal of bandwidth $k$. 

Recently, Tremblay et. al.~\cite{tremblayAB17} proposed the use of \textit{determinantal point processes (DPP)} in order to obtain the matrix ${\bf P}$ used in \cite{puysampling}. It is shown in~\cite{tremblayAB17} that an optimal ${\bf P}$ can be obtained
using DPP when ${\bf U}_{k}$ is known. Additionally, when the spectral decomposition is not accessible, it is shown how a variant of the Wilson's Algorithm introduced in~\cite{avenawilson} can be used in order to obtain a sampling set that can be shown 
is related with a DPP that leads to an approximate version of the optimal ${\bf P}$. The reconstruction of the signal is obtained by as solution of eqn.~(\ref{eq_puyreconst}); however, these results do not represent an improvement with respect to Anis
et. al.~\cite{ortega_proxies} or Chen et. al.~\cite{chensamplingongraphs} and may lead to larger reconstruction errors when the graph considered does not have a strong community graph structure~\cite{tremblayAB17}. Wang et. al.~\cite{7055883} consider the sampling 
and reconstruction of signals adapting concepts and ideas from frame theory, developing an iterative approach based on the concept of local sets. 

Marques et. al.~\cite{segarra_samlocagreg}, proposed a different approach with respect to previous works, considering the sampling and reconstruction of the signal using its samples on a single node. The central idea is based on the information provided
by the sequential application of the shift operator. The technique itself represents a novel alternative with potential applications in network analysis and its computational cost may be a drawback when large size graphs are considered.

\section{Blue-noise Sampling on Graphs}
This work proposes a different approach to graph signal sampling: the application of spatial dithering to the graph vertex domain where the spectral properties of well formed sampling patterns will equally benefit the graph vertex domain
as they do the spatial. This approach is motivated by the well established research in digital halftoning, which is the process of converting a continuous tone image or photograph into a pattern of printed
and not-printed dots for reproduction by inkjet or laser printers~\cite{ulichney,bluenoiseLaumagazine,BookLau}. Halftoning algorithms based on error-diffusion are of particular importance because they produce random patterns of homogeneously distributed dots where minority
pixels (black dots in highlights or white dots in shadows) are spaced as far apart as possible.  These patterns have power spectra dominated by high frequency energy, earning the name, ``blue-noise,'' since blue is the high frequency component
of white light.  Low frequency energy or red-noise contributes to halftone patterns looking fuzzy or noisy to the human visual system and are, therefore, to be avoided~\cite{ulichney,BookLau}.

%%%%%%%%%%%%%%%%%%%%%%%%%%%%%%%%%%%%%%%%%%%%%%%%
%\begin{figure}
%\centerline{\includegraphics[scale=1]{./figures/fig11}}
%\caption{Sampling and reconstruction showing (left) the power spectrum of the blue-noise sampling grid, (center) the circular band-limited surface to be sampled, and (right) spectral representation of the sampled signal with the ideal reconstruction 
%filter illustrated by the dashed line.}
%\vspace{-0.2in}
%\label{fig02}
%\end{figure}
%%%%%%%%%%%%%%%%%%%%%%%%%%%%%%%%%%%%%%%%%%%%%%%%

\begin{figure}[t!]
\centerline{\includegraphics[scale=0.08,angle=0]{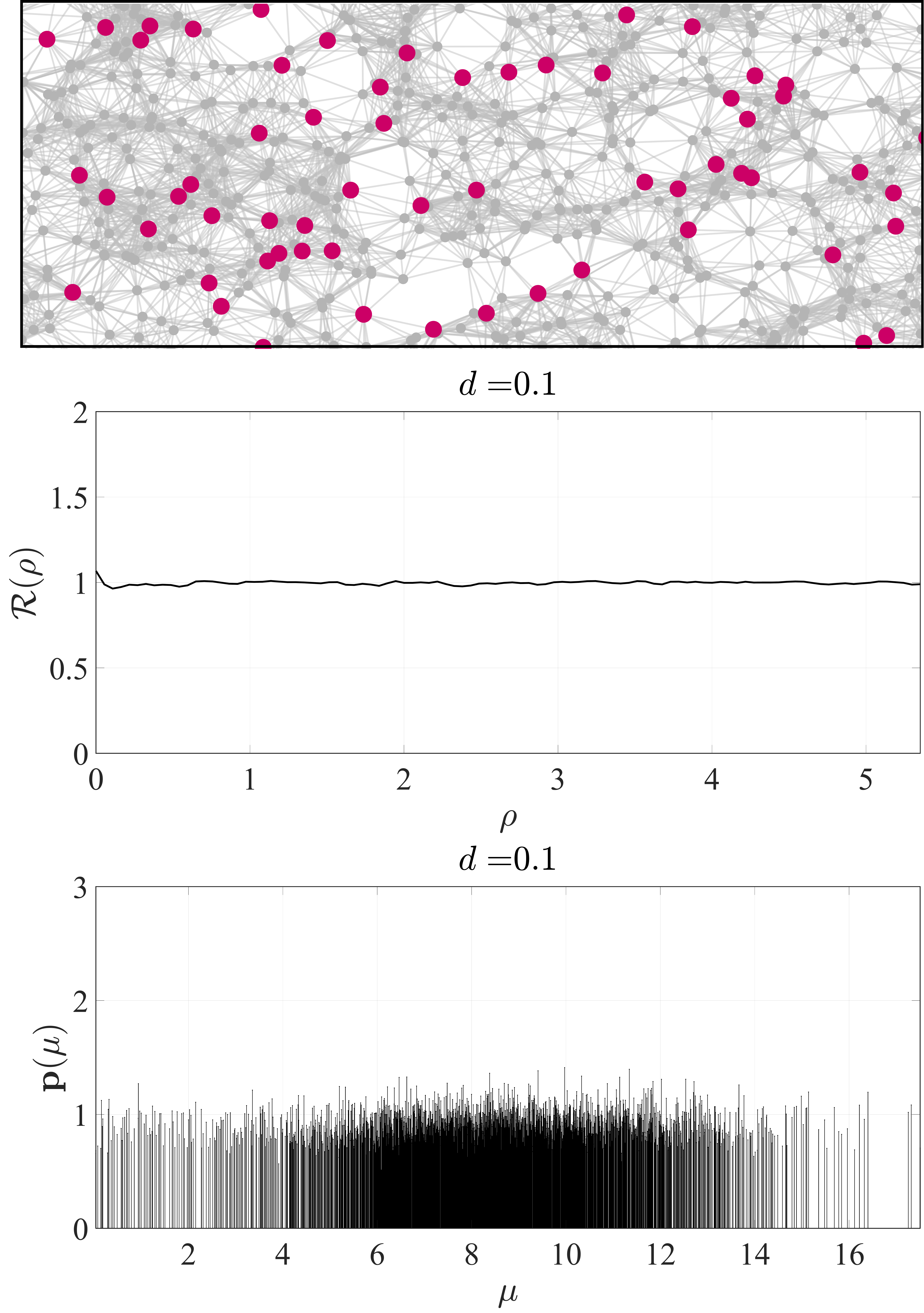}}
\caption{Illustration of the spatial and spectral properties of (top) a white-noise dither pattern on a Sensor Network graph with density, $d=0.1$, with (center) a flat pair correlation approximately equal to 1.0 for all internode distances, $\rho$, and (bottom) an approximately flat 
power spectra for all frequencies, $\mu$.}
\vspace{-0.2in}
\label{fig_sampwn}
\end{figure}

In order to establish a blue-noise model for sampling signals on a graph, we first propose the idea of a binary dither pattern on a graph, $G=(V(G),E(G))$, as the binary graph signal, $\boldsymbol{s}\in\{0,1 \}^{N}$. We refer to the fraction of samples 
that we intend to preserve as the density $d=m/N$, where $\Vert\boldsymbol{s}\Vert_{0}=m$. In the case of a white-noise dither pattern as illustrated in Fig.~\ref{fig_sampwn}~(top) on a Sensor Network graph for $d=0.1$, $\boldsymbol{s}$ is selected  uniformly 
at random from the space of binary signals
for which $\Vert\boldsymbol{s}\Vert_{0}=dN$; therefore each component of $\boldsymbol{s}$ can be modeled as a Bernoulli random variable with expected value $\mathbb{E}\{ \boldsymbol{s}(\ell)\}=d$.

\subsection{Vertex-domain Characteristics}
%%%%%%%%%%%%%%%%%%%%%%%%%%%%%%%%%%%%%%%%%%%%%%%%%%%%%%%%%%%%%%%%%%%%%%%%%%%%%%%%%%%%%%%
%%%%%%%%%%%%%%%%%%%%%%%%%%%%%%%%%%%%%%%%%%%%%%%%%%%%%%%%%%%%%%%%%%%%%%%%%%%%%%%%%%%%%%%
We define blue-noise sampling on graphs in terms of its desired vertex domain characteristics which resemble the spatial characteristics of blue-noise sampling in traditional halftoning. As such, we need to define a measure of spacing between neighboring nodes on a graph by defining a path between the nodes $v_{a}$ and $v_{b}$ by the {\em sequence} $(v_{a},u_{1},u_{2},\ldots ,u_{n},v_{b})$ where each node in the sequence indicates the nodes visited when going from $v_{a}$ to $v_{b}$, visiting between nodes with edge weights that are different from zero. Having a sequence of nodes defining a path, we define the length of this path according to:
\begin{multline}
\vert(v_{a},u_{1},u_{2},\ldots ,u_{n},v_{b})\vert=\\
\mathbf{W}(v_{a},u_{1})+\mathbf{W}(u_{1},u_{2})+\cdots +\mathbf{W}(u_{n},v_{b}),
\label{eq_shortestpathgraph}
\end{multline}
where the shortest path between two nodes, $v_{a}$ and $v_{b}$, is the path with minimum length and is represented by $\gamma_{v_{a},v_{b}}$. For any $v\in V(G)$, the \textit{open ball} of radius $\rho$ and centered in $v$ is defined
as $B(v,\rho)=\{ u\in V(G): \vert\gamma_{v,u}\vert<\rho \}$. The symbol $\mathbf{\Gamma}\in\mathbb{R}^{N\times N}$ represents the matrix of geodesic distances in the graph, where $\mathbf{\Gamma}(u,v)=\vert \gamma_{u,v}\vert$. We will 
refer to a collection of subsets of $V(G)$ as a \textit{cover} if the union of such subsets is equal to $V(G)$, and the cover will be called disjoint if the subsets are pairwise disjoint. 

Having defined the notion of distance on the vertex domain of the graph, we can introduce blue-noise sampling taking into account its characteristics in traditional halftoning. Blue-noise halftoning is characterized on the spatial domain
by a distribution of binary pixels
where the minority pixels are spread as homogeneously as possible. Distributing pixels in this manner creates a pattern that is aperiodic, isotropic (radially symmetric), and does not
contain any low-frequency spectral components. Halftoning a continuous-tone, discrete-space, monochrome image with blue-noise produces a pattern that, as Ulichney~\cite{ulichney} describes, is visually ``pleasant'' and ``does not clash with the structure of
an image by adding one of its own, or degrade it by being too `noisy' or uncorrelated.'' Similarly on a graph, the minority nodes composing the binary signal are expected to be equally spaced apart when measuring distance as the sum of the weights forming
the shortest path. With these ideas, we formally introduce blue-noise in the following definition. 

\begin{definition}[Blue-Noise on Graphs]
	Let $S\subset V(G)$ be a subset of nodes in the graph $G$ with $S=\{ s_{1},s_{2},\ldots s_{m}  \}$. Then, it is said that $S$
	represents an ideal blue-noise sampling pattern, if the following conditions are satisfied:
	\begin{itemize}
		\item There is a collection of open balls $B(s_{i},\lambda)$ that forms a cover of $V(G)$.
		\item The value of $\lambda$ is the minimum possible for all the subsets of nodes of size $m$.
	\end{itemize}
	\label{def_BN}
\end{definition}
\noindent Definition \ref{def_BN} implies that ideal blue-noise sampling patterns have their sampling nodes located as far as possible from each other, or in other words, there is a typical vertex domain spreading of the sampling nodes. Figure~\ref{fig_sampbn}~(top) 
illustrates a typical blue-noise pattern on a sensor network. We use this attribute as the defining characteristic of a blue-noise sampling pattern; however, we will show in later sections that, in some classes of graphs this vertex domain spreading implies or 
is correlated with a high frequency behavior on the spectral domain.

\subsubsection{Vertex-domain Metrics}
\begin{figure}
\centering\includegraphics[scale=0.6,angle=-90]{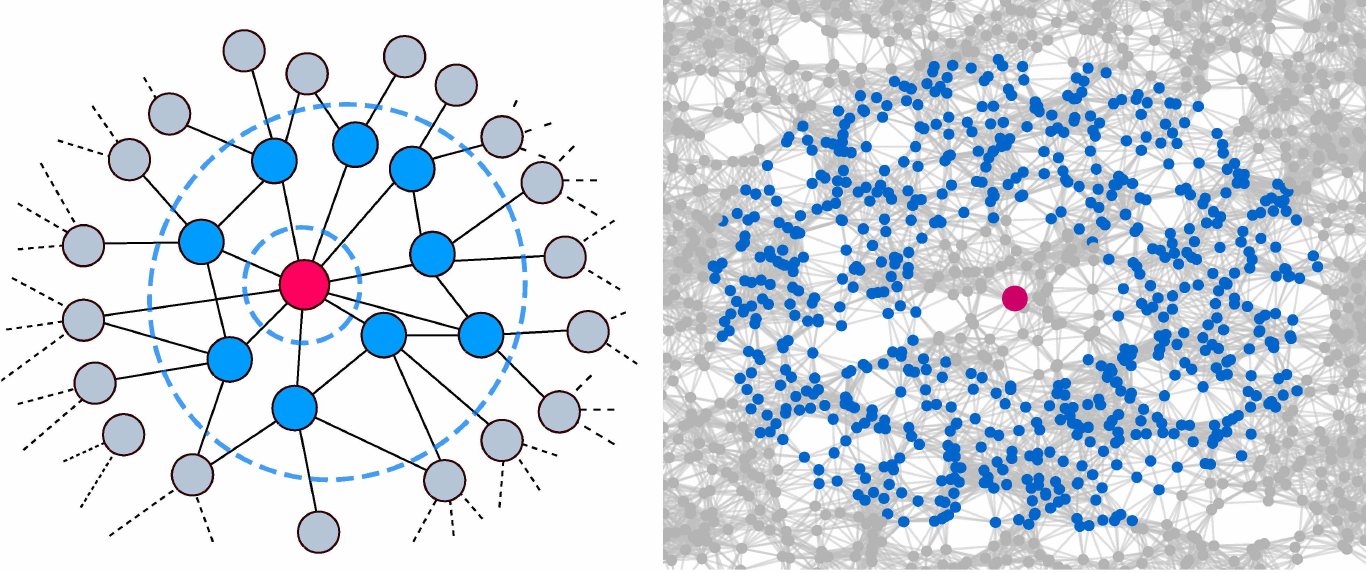}
\caption{Illustration of $B_{\theta}(v,\rho)$ in a graph. Left: representation of $B_{\theta}(v,\rho)$ for small values of $\rho$ and $\theta$. Right: Illustration of $B_{\theta}(v,\rho)$ for large values of $\rho$ and $\theta$.
The nodes in blue color are located in the annulus of radius $\rho$ and width $\theta$ centered at the node $v$ indicated in red color.}
\vspace{-0.1in}
\label{fig_Bneightongraph}
\end{figure}

For any $v\in V(G)$, the \textit{annulus} of radius $\rho$, width $\theta$, and center $v$ is defined as $B_{\theta}(v,\rho)=\{ u\in V(G): \rho-\theta\leq\vert\gamma_{v,u}\vert<\rho+\theta \}$. Figure~\ref{fig_Bneightongraph} illustrates
an example of $B_{\theta}(v,\rho)$. 
%%%%%%%%%%%%%%%%%%%%%%%%%%%%%%%%%%%%%%%%%%%%%%%%%%%%%%%%%%%%%%%%%%%%%%%%%%%%%%%%%%%%%%%
%%%%%%%%%%%%%%%%%%%%%%%%%%%%%%%%%%%%%%%%%%%%%%%%%%%%%%%%%%%%%%%%%%%%%%%%%%%%%%%%%%%%%%%
With a notion of concentric rings in $B_{\theta}(v,\rho)$, we can now define the pair correlation on a graph. Specifically, let $S=\supp (\boldsymbol{s})=\{ s_{1}, s_{2}, \ldots, s_{m}\}$ be the support of the sampling
pattern $\boldsymbol{s}$ and let $\left\Vert \boldsymbol{s}(B_{\theta}(s_{i},\rho))\right\Vert_{0}$ be the number of 1s of  $\boldsymbol{s}$ on 
$B_{\theta}(s_{i},\rho)$, then the sample pair correlation function, $\mathcal{R}_{\boldsymbol{s}}(\rho)$, associated to $\boldsymbol{s}$ is defined by 
\begin{equation}
\mathcal{R}_{\boldsymbol{s}}(\rho)=\frac{\frac{1}{m} \sum\limits_{i=1}^{m} \left\Vert \boldsymbol{s}(B_{\theta}(s_{i},\rho))\right\Vert_{0}}
{\frac{1}{N}\sum\limits_{v\in V(G)}\Vert   \boldsymbol{s}(B_{\theta}(v,\rho) )  \Vert_{0} }.
\label{eq_samplepaircorr}
\end{equation}
 Notice that the numerator in (\ref{eq_samplepaircorr}) indicates the average number of 1s in $\boldsymbol{s}$ on a ring of width $\theta$ that is centered on a 1 of $\boldsymbol{s}$, while the denominator indicates the average number of 1s on the ring of 
the same width when it is 
centered at any arbitrary node. Now, the pair correlation for $q$ realizations $\boldsymbol{s}_{1},\ldots,\boldsymbol{s}_{q}$ of a random sampling pattern is defined as
\begin{equation}
\mathcal{R}(\rho)=\frac{1}{q}\sum_{r=1}^{q}\mathcal{R}_{\boldsymbol{s}_{r}}(\rho),
\label{eq_paircorr}
\end{equation}
as the influence of a sampling point at node $v$ on all other nodes in the geodesic annular region $B_{\theta}(v,\rho)$.  Notice that for the computation of eqn.~(\ref{eq_samplepaircorr}) several values of $\theta$ can be considered, in this work the value of $\theta$ is the average of nonzero edge weights. 

Note that a maxima
of $\mathcal{R}(\rho)$ can be considered as an indication of the frequent occurrence of the inter-node distance, $\rho$, between nodes set to 1 whereas minima indicate a reduced occurrence.  Since for random patterns the expected number 
of 1s in any annular ring is proportional to the number of nodes within the ring, we expect a pair correlation equal to 1 for all $\rho > 0$ as illustrated in Fig.~\ref{fig_sampwn}~(center).

Blue-noise, when applied to an image of constant gray-level $g$, spreads the minority pixels of the resulting binary image as homogeneously as possible such that the pixels are separated by an average distance, $\lambda_b$, referred to as the {\em principal wavelength} of blue-noise. These minority pixels are the pixels that are used to represent 
the properties of a region on a grayscale image, for instance in a dark region the minority pixels are labeled with 1 while in white regions the minority pixels are labeled with 0. Then, the value of $\lambda_b$ is defined as the radius of a round disc, surrounding a minority pixel, such that the ratio of the surface area of a minority pixel to the surface area of the disc is equal to the density of minority pixels, $d=g$ for $0<g\leq1/2$ and $d = 1-g$ for $1/2<g\leq1$, which we can write as:
%%%%%%%%%%%%%%%%%%%%%%%%%%%%%%%%%%%%%%%%%%%%%%%
\begin{equation}
d = \frac{D_x D_y}{\lambda_b^2},
\label{blue_noise_principal_wavelength_eqn}
\end{equation}
%%%%%%%%%%%%%%%%%%%%%%%%%%%%%%%%%%%%%%%%%%%%%%%
where $D_x$ and $D_y$ are the sampling periods (distance between samples) of the digital image in the $x$ and $y$ directions, respectively.

\begin{figure}
	\centering
	\includegraphics[scale=0.315,angle=-90]{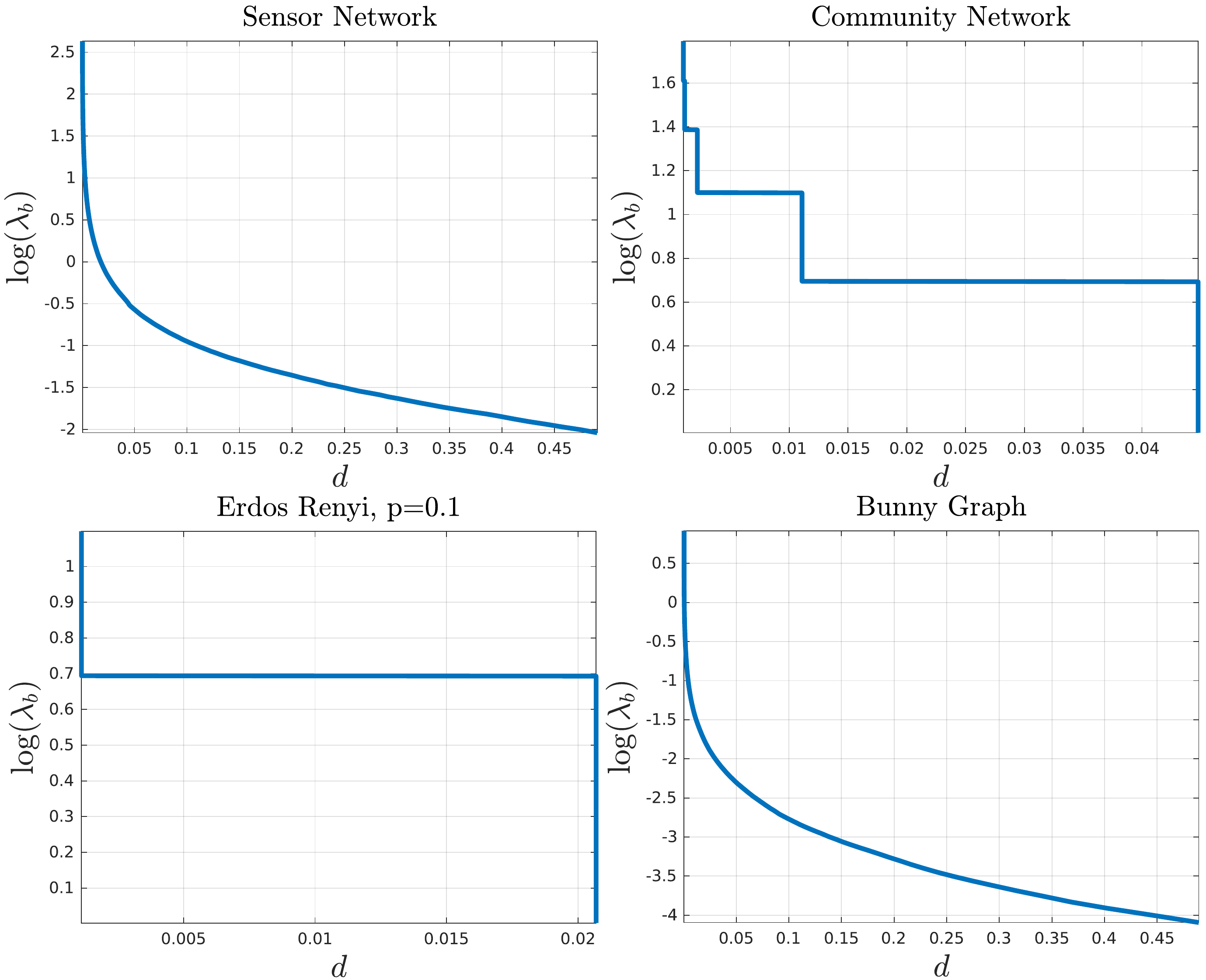}
	\caption{Principal wavelength $\lambda_{b}$  versus the density of the sampling pattern for  four different graphs traditionally considered in the literature. }
	\label{fig_lambdabvsd}
\end{figure}

In order to extend the notion of principal wavelength to graphs, we need a notion of surface area as the expected number of graph nodes, $\mathbb{E}\{ {\cal N}(\lambda) \}$, within a distance or path length, $\lambda$, of a given minority node.
We expect the ratio of our single, minority node to all nodes within a path length, $\lambda_b$, to equal the density level according to:
%%%%%%%%%%%%%%%%%%%%%%%%%%%%%%%%%%%%%%%%%%%%%%%
\begin{equation}
d = \frac{1}{\mathbb{E}\{ {\cal N}(\lambda_b) \}}.
\label{graph_blue_noise_principal_wavelength_eqn}
\end{equation}
%%%%%%%%%%%%%%%%%%%%%%%%%%%%%%%%%%%%%%%%%%%%%%%
Being that $\mathbb{E}\{ {\cal N}(\lambda_b) \}$ is graph dependent, the graph blue-noise wavelength, $\lambda_b$, is likewise graph dependent and its characterization is still an open area of research~\cite{vdsilva_ghrist1,desilva2007,Silva07homologicalsensor}. In general, one can derive $\lambda_b$ versus $d$ experimentally as we have in Fig.~\ref{fig_lambdabvsd} where we show the principal wavelength versus the density sampling $d$ for some commonly used graphs.  We note that in the case of the sensor graph, $\lambda$ varies smoothly with $d=1/\mathbb{E}\{ {\cal N}(\lambda_b) \}$ while, in the case of the community graph, it varies with a piecewise constant behavior with respect to $d$.

%%%%%%%%%%%%%%%%%%%%%%%%%%%%%
%%%%%%%%%%%%%%%%%%%%%%%%%%%%

\begin{figure}
	\centerline{\includegraphics[width=3.5in]{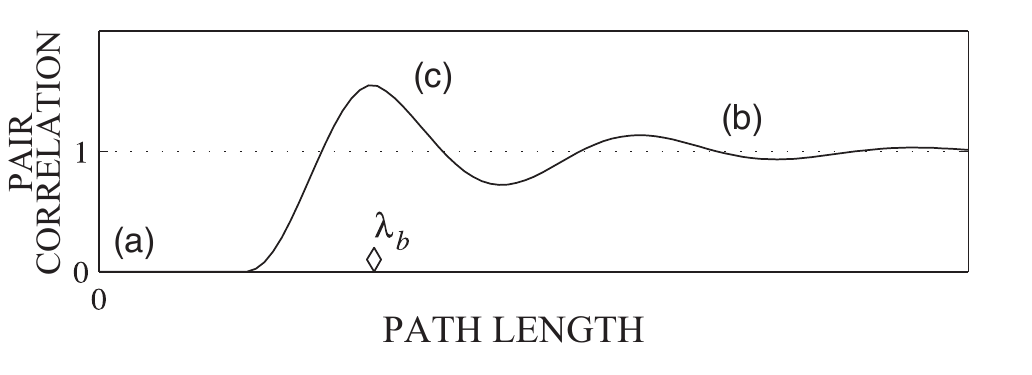}}
	\centerline{\includegraphics[width=3.5in]{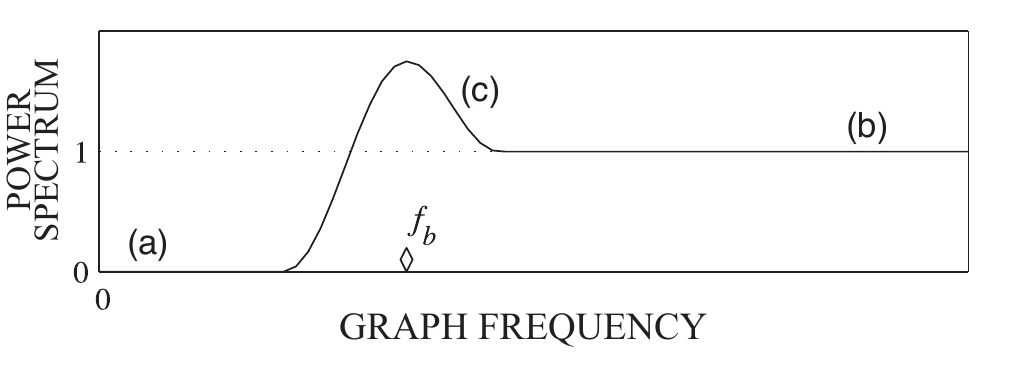}} 
	\caption{The ideal (top) pair correlation and (bottom) power spectra for blue-noise sampling patterns.}
	\label{fig10}
\end{figure}

In light of the nature of graph blue-noise to isolate minority nodes, we can begin to characterize blue-noise graph signals in terms of the pair correlation, ${\cal R}(\rho)$, by noting that: (a) few or no neighboring minority nodes lie within a path
length of $\rho<\lambda_{b}$; (b) for $\rho>\lambda_{b}$, the expected number of minority nodes per unit area tends to stabilize around a constant value; and (c) the average number of minority nodes within the path length, $\rho$, increases sharply
nearly $\lambda_{b}$. The resulting pair correlation for blue-noise is, therefore, of the form in Fig.~\ref{fig10}~(top), where ${\cal R}(\rho)$ shows: (a) a strong inhibition of minority nodes near $\rho=0$, (b) a decreasing correlation of minority
nodes with increasing $\rho$ ($\lim_{\rho\rightarrow \infty}{\cal R}(\rho)=1$), and (c) a frequent occurrence of the inter-node distance $\lambda_b$, the principal wavelength, indicated by a series of peaks at integer multiples of $\lambda_b$.  The principal
wavelength is indicated in Fig.~\ref{fig10}~(top) by a diamond located along the horizontal axis. Returning to the sample blue-noise signal of Fig.~\ref{fig_sampbn}~(top), the resulting pair correlation of Fig.~\ref{fig_sampbn}~(center) has a principal
wavelength of $\lambda_{b} = 0.56$ with a clearly visible peak of 1.55, meaning that nodes equal to 1 are $55\%$ more likely to occur at a distance of $\rho = 0.56$ from an existing 1 than for the unconstrained probability of a node being equal to 1. 

%%%%%%%%%%%%%%%%%%%%%%%%%%%%%%%%%%%%%%%%%%%%%%%

\subsection{Spectral Characteristics}
%%%%%%%%%%%%%%%%%%%%%%%%%%%%%%%%%%%%%%%%%%%%%%%%%%%%%%%%%%%%%%%%%%%%%%%%%%%%%%%%%%%%%%%
\begin{figure}[t!]
	\centerline{\includegraphics[scale=0.08,angle=0]{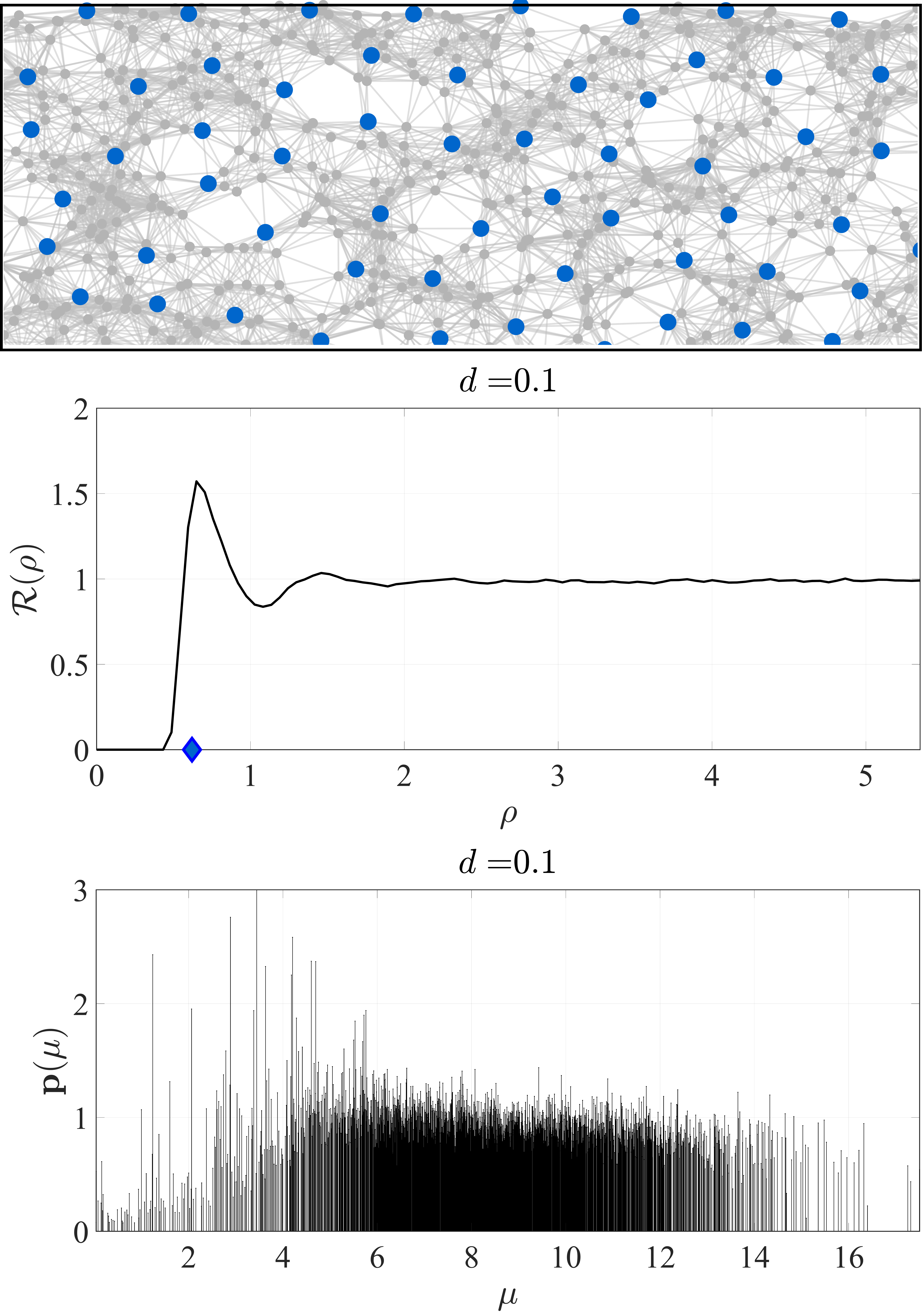}}
	\caption{Illustration of the spatial and spectral properties of (top) a blue-noise dither pattern on a Sensor Network graph with density, $d=0.1$, with (center) a pair correlation peak at the principal wavelength, $\lambda_b$, and (bottom) an approximately 
		high frequency only power spectrum for frequencies, $\mu$.}
	\label{fig_sampbn}
\end{figure}
%%%%%%%%%%%%%%%%%%%%%%%%%%%%%%%%%%%%%%%%%%%%%%%%%%%%%%%%%%%%%%%%%%%%%%%%%%%%%%%%%%%%%%%
Blue noise sampling patterns are characterized in traditional halftoning for a high frequency behavior~\cite{bluenoiseLaumagazine}. In this section we state a connection between the spectral characteristics of a sampling pattern and its vertex domain 
characteristics, using the local properties of partitions of $V(G)$ that are measured by the isoperimetric constants of local induced subgraphs. In order to characterize the frequency content of a sampling pattern, a cost function 
is proposed. In particular,
we propose a scalar measure of low-frequency energy, $R_{\boldsymbol{s}}$, in the signal $\boldsymbol{s}$, as the weighted sum of all Fourier coefficients' energies:
\begin{equation}
R_{\boldsymbol{s}}= \frac{1}{\Vert\hat{\boldsymbol{s}}\Vert_{2}^{2}}\sum_{\ell=2}^{N}\frac{\hat{\boldsymbol{s}}^{2}(\ell)}{\mu_{\ell}}=\frac{1}{m}\sum_{\ell=2}^{N}\frac{\hat{\boldsymbol{s}}^{2}(\ell)}{\mu_{\ell}},
\label{eq_bluenoisedef}
\end{equation}
where $\hat{\boldsymbol{s}}$ is the graph Fourier transform of $\boldsymbol{s}$. $R_{\boldsymbol{s}}$ is coined as the \textit{redness} of $\boldsymbol{s}$ as it measures low frequency spectral content.

In order to establish a connection between $R_{\boldsymbol{s}}$ and the vertex domain characteristics of a sampling pattern, it is important to consider the following theorems. 
\begin{theorem}
	For the graph $G=(V(G),E(G))$, let $\mathcal{P}=\left\lbrace V(\Omega_{1}),V(\Omega_{2}),\ldots,V(\Omega_{\vert \mathcal{P}\vert})\right\rbrace$ be a partition of $V(G)$, where $\Omega_{j}$ is the induced subgraph given by 
	$V(\Omega_{j})$.  Let $\delta_{j}$ be the isoperimetric dimension of $\Omega_{j}$. Then if
	\begin{equation}
	\delta_{1}=\delta_{2}=\ldots =\delta_{\vert\mathcal{P}\vert}=\delta
	\end{equation}
	it follows that
	\begin{equation}
	\Lambda_{\mathcal{P}}>\min\left\lbrace  
	C_{\delta}\left( \frac{1}{vol(\Omega_{1})}\right)^{\frac{2}{\delta}},\ldots , C_{\delta}\left( \frac{1}{vol(\Omega_{\vert \mathcal{P}\vert})}\right)^{\frac{2}{\delta}}
	\right\rbrace
	\label{eq_Lambdapartitions1}
	\end{equation}
	where $C_{\delta}$ is a constant that depends on $\delta$.\\
	Proof: See Appendix \ref{appendix_prooftheorem_Lambdapartitions1}
	\label{theorem_Lambdapartitions1}
\end{theorem}
\noindent Theorem~\ref{theorem_Lambdapartitions1} indicates that when the graph has a local invariant isoperimetric dimension, the quality of a partition $\mathcal{P}$ for the representation of bandlimited signals, measured by $\Lambda_{\mathcal{P}}$, is defined by the set in $\mathcal{P}$ with the largest volume. The
concept of isoperimetric dimension, originally defined on manifolds, provides a measure of
how similar is the global behavior of 
a manifold with respect to a Euclidean space~\cite{chungisoperimetric}. Similarly,  in the case of graphs, the isoperimetric dimension indicates how close the behavior of a graph is with respect to regular grid-like graphs. For instance, the 
isoperimetric dimension of 
the $n-$dimensional regular grid is $n$~\cite{chungisoperimetric}.  In the following theorem, we indicate when the right-hand side of eqn.~(\ref{eq_Lambdapartitions1}) is maximized. 

%%%%%%%%%%%%%%%%%%%%%%%%%%%%%%%%%%%%%%%%%%%%%%%%%%%%%%%%%%%%%%%%%%%%%%%%%%%%%%%%%%%%%%%
\begin{figure}[t!]
	\centerline{\includegraphics[scale=1,angle=-90]{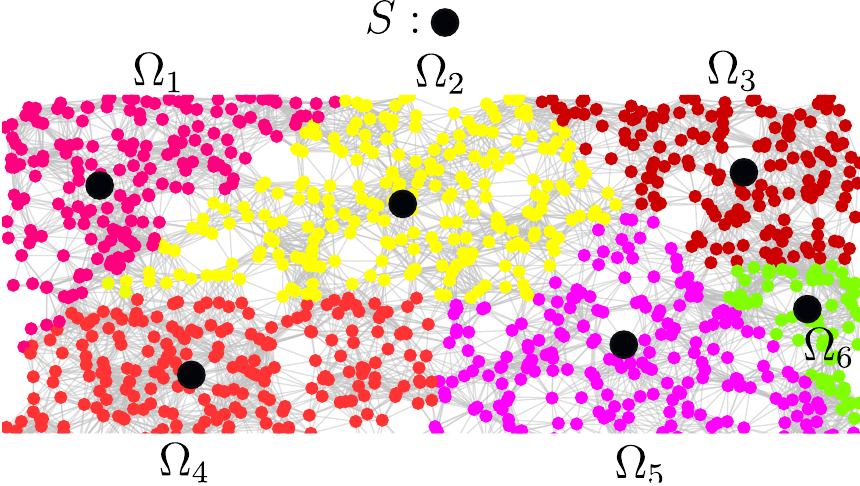}}
	\caption{Partition of $V(G)$ for a graph $G$. Each color indicates the subgraph $\Omega_{j}$ induced by $V(\Omega_{j})$.  Illustration shows how a sampling pattern can be built from the partition selecting the
		sampling nodes on the set $S$, whose nodes are indicated in black color. Notice that the sampling pattern indicated satisfies eqn.~(\ref{eq_homogcond1}) and eqn.~(\ref{eq_homogcond2}).}
	\label{fig_partandS}
\end{figure}
%%%%%%%%%%%%%%%%%%%%%%%%%%%%%%%%%%%%%%%%%%%%%%%%%%%%%%%%%%%%%%%%%%%%%%%%%%%%%%%%%%%%%%%
\begin{theorem}
Under the conditions stated in Theorem~\ref{theorem_Lambdapartitions1} and 
for a fixed value of $\vert \mathcal{P}\vert$, the partition that maximizes the right hand side of eqn.~(\ref{eq_Lambdapartitions1}) satisfies that
\begin{equation}
 vol(\Omega_{i})=vol(\Omega_{j}) \quad\forall i,j.
\end{equation}
Proof: See Appendix \ref{appendix_prooftheorem_Lambdapartitions2}
\label{theorem_Lambdapartitions2}
\end{theorem}
\noindent Under the conditions stated in Theorem~\ref{theorem_Lambdapartitions1}, Theorem~\ref{theorem_Lambdapartitions2} provides the characteristics of the partition that will maximize the bandwidth of signals that can be represented in a unique way via 
their average values on the elements of the partition.

Now, it is important to notice that for any partition $\mathcal{P}=\left\lbrace V(\Omega_{1}), V(\Omega_{2}),\ldots ,V(\Omega_{\vert \mathcal{P}\vert})\right\rbrace$, it is possible to build a sampling pattern, locating one sampling node per 
partition element (see Fig.~\ref{fig_partandS}). In the following theorem, we show that the spectral characteristics of such sampling patterns, measured by $R_{\boldsymbol{s}}$, are bounded by the local characteristics of the elements in $\mathcal{P}$.

\begin{theorem}
	Let $\mathcal{P}=\left\lbrace V(\Omega_{1}), V(\Omega_{2}),\ldots ,V(\Omega_{\vert \mathcal{P}\vert})\right\rbrace$ a partition of $V(G)$  and 
	let 
	$\boldsymbol{s}\in\{0,1\}^{N}$ a sampling pattern chosen according to
	\begin{equation}
	\Vert \boldsymbol{s}(V(\Omega_{j}))\Vert_{0}=1\quad\forall j
	\label{eq_homogcond1}
	\end{equation}
	\begin{equation}
   \text{If}~\boldsymbol{s}(v)=1,~\text{then}~\boldsymbol{s}(u)=0\quad\forall u\sim v
	\label{eq_homogcond2}
	\end{equation}
	then
		\begin{equation}
		R_{\boldsymbol{s}}\leq\frac{(\mu_{2}+\mu_{N})^{2}(1-\vert\mathcal{P}\vert/N)^{2}
			}{4\mu_{2}\mu_{N}\min\limits_{j}\left\lbrace \frac{C_{\delta_{j}}}{vol(\Omega_{j})^{2/\delta_{j}}}\right\rbrace}.
		\label{eq_spvsred1}
		\end{equation}
	If in addition, $\delta=\delta_{1}=\ldots=\delta_{\vert\mathcal{P}\vert}$ and $vol(\Omega)=vol(\Omega_{1})=\ldots =vol(\Omega_{\vert\mathcal{P}\vert})$, then
	\begin{equation}
	R_{\boldsymbol{s}}\leq\frac{(\mu_{2}+\mu_{N})^{2}(1-\vert\mathcal{P}\vert/N)^{2}vol(\Omega)^{\frac{2}{\delta}}}{4C_{\delta}\mu_{2}\mu_{N}}.
	\label{eq_spvsred2}
	\end{equation}	
	Proof: See Appendix \ref{appendix_theorem_parsandsamppattern}
	\label{theorem_parsandsamppattern}
\end{theorem}
\noindent In order to discuss the meaning and implications of Theorem~\ref{theorem_parsandsamppattern}, it is important to mention that eqn.~(\ref{eq_homogcond1}) and eqn.~(\ref{eq_homogcond2}) imply that 
there is one sampling node per element of the partition with $\Vert\boldsymbol{s}\Vert_{0}=\vert\mathcal{P}\vert$, 
and that there is a minimum interdistance between the sampling nodes in $\boldsymbol{s}$ (see Fig.~\ref{fig_partandS}). In particular, eqn.~(\ref{eq_homogcond2}) assures that the sampling points in $\boldsymbol{s}$ are far from the boundaries 
of the elements of the partition. 

Notice that eqn.~(\ref{eq_spvsred1}) presents a general upper bound for the redness of an arbitrary sampling pattern subject to eqn.~(\ref{eq_homogcond1}) and eqn.~(\ref{eq_homogcond2}). Meanwhile, eqn.~(\ref{eq_spvsred2}) provides a tighter bound 
that is connected with blue-noise sampling patterns as a consequence of having  
the elements in the partition with the same volume and the same isoperimetric dimension. In this second case, we see that as the size  of the partition, $\mathcal{P}$, increases (and therefore the number of sampling nodes in $\boldsymbol{s}$) $vol(\Omega)$
decreases and so it is the value $R_{\boldsymbol{s}}$, making clear the connection between a uniform vertex spreading of the sampling nodes in $\boldsymbol{s}$ and a low redness. As a consequence, a behavior like the one depicted in Fig.~\ref{fig_sampbn}~(bottom) 
is expected. It is important to emphasize that this statement is connected to Theorems~\ref{theorem_Lambdapartitions1} and~\ref{theorem_Lambdapartitions2}, where the characteristics of good partitions for the representation of bandlimited signals is stated. In the case of the traditional halftoning scenario where the problem is modeled on a 2-dimensional grid, the
conditions of Theorems~\ref{theorem_parsandsamppattern}, \ref{theorem_Lambdapartitions2} and~\ref{theorem_Lambdapartitions1} hold and a typical response like the one shown in Fig.~\ref{fig10}~(bottom) is obtained. 

Theorem~\ref{theorem_parsandsamppattern} also implies that there is an upper limit about the number of elements in the partition $\mathcal{P}$ that can be considered, and with that, it comes a limitation in the number of sampling nodes for which these inequalities
hold. In particular, for very large values of $\vert \mathcal{P}\vert$,  eqn.~(\ref{eq_homogcond1}) and eqn.~(\ref{eq_homogcond2}) cannot be satisfied.  We point out that this does not diminish the quality of the sampling patterns, but 
instead points out that the relationship between the spectral domain and the vertex domain is not guaranteed to be governed by eqn.~(\ref{eq_spvsred1}).

\subsubsection{Spectral Metrics}
It is also possible to characterize the spectral properties of binary dither patterns on a graph where we extend the idea of periodograms to graphs such that the GFTs of $q$ realizations of $\boldsymbol{x}$, i.e.~$\boldsymbol{x}_{1},\boldsymbol{x}_{2},\ldots,\boldsymbol{x}_{q}$, are averaged together to form the power spectrum:
\begin{equation}
	\mathbf{p}(\ell)=\frac{N}{q}\sum_{i=1}^{q}\frac{\hat{\boldsymbol{x}}_{i}(\ell)^{2}}{\Vert \hat{\boldsymbol{x}}_{i}\Vert_{2}^{2}}\quad \ell=2,\ldots,N.
	\label{eq_psdgraph}
\end{equation}
Notice that the $\ell^{th}$ component of $\mathbf{p}$ is associated with the $\ell^{th}$ eigenvalue $\mu_{\ell}$.
Like its binary halftone counterpart, the GFT of a white-noise sampling pattern is expected to be flat for all $\mu_k$s, and to visualize this power spectra, Fig.~\ref{fig_sampwn}~(bottom) shows an estimate of the power spectra for 100 unique white-noise dither patterns generated on the 2000-node Sensor Network graph with pattern density $d=0.1$.

\subsection{Blue-noise Sampling Sets}
In the following corollary we state how from a blue-noise sampling pattern a good partition in the sense of Theorem~\ref{pesentheorem2019_3} can be obtained.
\begin{corollary}
	Let $\boldsymbol{s}$ be a blue-noise sampling pattern obtained according to Definition \ref{def_BN}, with $\Vert\boldsymbol{s}\Vert_{0}=m$ and $\supp (\boldsymbol{s})=\{ s_{1}, s_{2}, \ldots , s_{m}\}$. Let $B(s_{j},\lambda)$ be as specified in Definition  \ref{def_BN}. Then, there exists a partition 
	$\mathcal{P}=\left\lbrace V(\Omega_{1}), V(\Omega_{2}),\ldots ,V(\Omega_{\vert \mathcal{P}\vert})\right\rbrace$ of $V(G)$ such that
	\begin{equation}
	u\in V(\Omega_{j})\Leftrightarrow u\in B(s_{j},\lambda), u\notin B(s_{i},\lambda)~\forall~i\neq j
	\end{equation}
	and the elements in the intersection between the sets $B(s_{i},\lambda)$ are distributed on the $V(\Omega_{i})$  such that the quantity $\sum_{i\neq j}\left\vert vol(\Omega_{i})-vol(\Omega_{j})\right\vert$
%	\begin{equation}
%	\sum_{i\neq j}\left\vert vol(\Omega_{i})-vol(\Omega_{j})\right\vert
%	\end{equation}
	is minimized. Additionally if $\Lambda_{\mathcal{P}}>(1+1/\alpha)\omega$, $\alpha>0$ any $\boldsymbol{x}\in PW_{\omega}(G)$ can be uniquely determined from  its values at $\{ s_{1}, s_{2}, \ldots , s_{m}\}$ always that 
	$\boldsymbol{x}(s_{j})=\boldsymbol{x}^{\mathsf{T}}(\boldsymbol{\xi}_{j}\circ\boldsymbol{\xi}_{j})~\quad\forall j$.	
	
	%\label{corollary_ssamptopart}
	Proof: See Appendix \ref{appendix_theorem_partitionfrompattern}.
	\label{theorem_partitionfrompattern}
\end{corollary}
\noindent This corollary indicates that given a sampling pattern whose sampling points are located as far as possible from each other, it is possible  to build a partition from which a unique representation of a set of bandlimited signals is possible. Additonally, if the conditions of Theorem~\ref{theorem_Lambdapartitions1} are satisfied, then the partitions obtained 
are the ones that maximize the value of 
$\Lambda_{\mathcal{P}}$. 

 Theorem~\ref{theorem_pensensonuniquesets} tells us that when a fixed value of the bandwidth $\omega$ is considered and a signal has to be sampled taking $m$ samples, it is necessary to look for the set of
nodes, $S$, such that $S^{c}$ is a $\Lambda_{S^{c}}$-removable set with $\omega<\Lambda_{S^{c}}$. Finding the subset of $m$ nodes with the maximum $\Lambda_{S^{c}}$ would, therefore, give 
the \textit{best sampling set}. On the other hand, if a signal has to be sampled taking a number of $m$ samples, choosing a set of nodes $S$ with the maximum value of $\Lambda_{S^{c}}$ will extend the class of signals,
$PW_{\omega}(G)$, that can be sampled and represented in a unique way with $m$ samples. 

Now if one can show that minimizing the redness in a sampling signal promotes high values of  $\Lambda_{S^{c}}$, one could argue blue-noise was a desirable attribute for efficient sampling. The following theorem establishes
this relationship:
\begin{theorem}
Let $\boldsymbol{s}:V(G)\longrightarrow \{0,1\}$ be a sampling pattern with $\boldsymbol{s}({S})=1$, $\boldsymbol{s}({S^{c}})=0$ for $S\subset V(G)$ and
$\vert S\vert=\Vert\boldsymbol{s}\Vert_{0}=m$, then the $\Lambda_{S^{c}}-$constant of the set $S^{c}$ satisfies 
\begin{equation}
\Lambda_{S^{c}}>C_{\delta}\left( \frac{R_{\boldsymbol{s}}}{vol(G)R_{\boldsymbol{s}}-m\left(1-\frac{m}{N}\right)^{2}}\right)^{\frac{2}{\delta}} 
\label{eq_theorem_lambdaremov_bncostfunc1}
\end{equation}
where $R_{\boldsymbol{s}}$ is the redness in $\boldsymbol{s}$ from eqn.~(\ref{eq_bluenoisedef}); 
%\begin{equation}
%vol(G)=\sum_{v=1}^{N}\mathbf{D}(v,v);
%\end{equation}
$\delta$ is the isoperimetric dimension of $G$~\cite{fuhrpesenson,chung1997spectral}; and $C_{\delta}$ a constant that depends only on $\delta$.\\
Proof: See Appendix \ref{appendix_theorem_lambdaremov_bncostfunc}.
\label{theorem_lambdaremov_bncostfunc}
\end{theorem} 
\noindent To summarize, Theorem~\ref{theorem_lambdaremov_bncostfunc} tells us that the best sampling set, $S$, is the one for which the value of $\Lambda_{S^{c}}$ is a maximum; therefore while blue-noise sampling patterns (which minimize $R_{\boldsymbol{s}}$) are not necessarily the {\em best} sampling sets, they are {\em good} sampling sets. Notice that eqn.~(\ref{eq_theorem_lambdaremov_bncostfunc1}) is well defined as $vol(G)R_{\boldsymbol{s}}-m\left(1-\frac{m}{N}\right)^{2}>0$, which is tight when $S^{c}\cup bS^{c} = V(G)$ where $bS^{c}$ is the boundary of $S^{c}$. This criteria can be satisfied making the nodes in $S$ as spread apart as possible in the graph, which is reasonable as a sampling set where all the nodes are too concentrated in one area could lead to poor reconstructions of signals that exhibit fast changes in the sparsely sampled areas left elsewhere. 

As an approach to reinforce the benefits of blue-noise sampling sets, we can use the quantities introduced in Theorem~\ref{theorem_Kstheory} to show how blue-noise promotes those sampling sets that maximize the bandwidth of the signals that can be represented in a unique way on a given sampling set as indicated in the following theorem:
\begin{theorem}
Let $\boldsymbol{s}:V(G)\longrightarrow \{0,1\}$ with $\boldsymbol{s}({S})=1$, $\boldsymbol{s}({S^{c}})=0$, $S\subset V(G)$. If $K_{S}>0$, then
\begin{equation}
K_{S}\geq \left(\frac{m\left( 1-\frac{m}{N} \right)^{2}}{R_{\boldsymbol{s}}}-\gamma \right)^{1/2} 
\end{equation} 
where $\gamma=\max_{S,v^{'},v}\left(\sum_{v\in S^{c}\setminus v^{'}}w_{S}(v)^{2}\right),$
%\begin{equation*}
%\gamma=\max_{S,v^{'},v}\left(\sum_{v\in S^{c}\setminus v^{'}}w_{S}(v)^{2}\right) .
%\end{equation*}
and $R_{\boldsymbol{s}}$ is, again, the redness in $\boldsymbol{s}$ from eqn.~(\ref{eq_bluenoisedef}).\\
Proof: See Appendix \ref{appendix_Kstheorem}.
\label{theorem_Ks}
\end{theorem}
\begin{figure*}[t!]
	\centering\includegraphics[scale=1,angle=0]{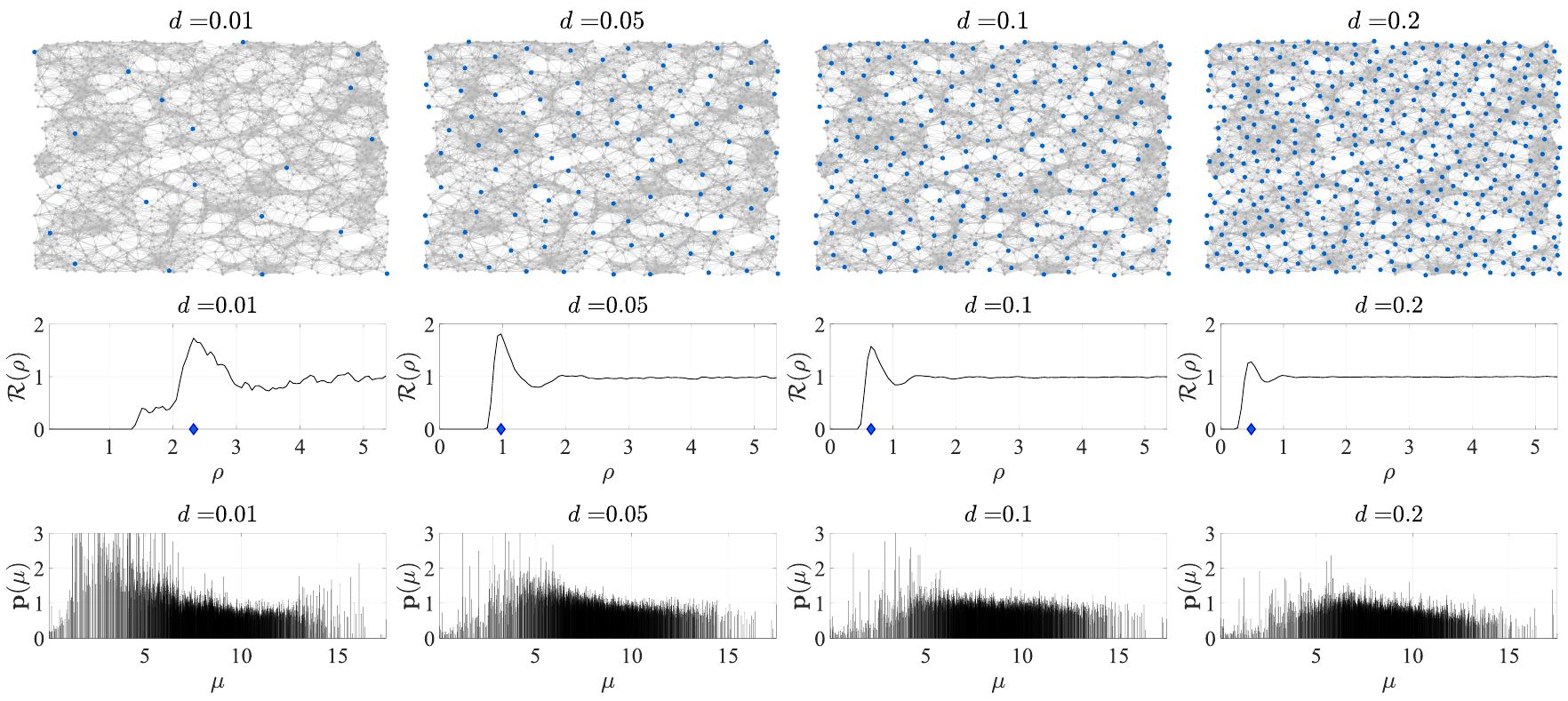} 
	\caption{Void and cluster blue-noise sampling patterns for different intensities $d$ for a sensor network graph. First row: Localization on the graph of the nodes selected in a blue-noise sampling pattern. Second row: The pair correlation function $\mathcal{R}(\rho)$ for the sampling patterns indicating with a diamond marker the value of $\lambda_{b}$. Third row: Power spectral density for the different blue-noise sampling patterns.}
	\label{fig_BNbigpicture2sensor}
\end{figure*}
\noindent Theorem~\ref{theorem_Ks} indicates that lowering the redness of the sampling pattern raises the minimum possible value of $K_{S}$ and, therefore, extends the set of signals, $PW_{\omega}(G)$, that can be represented in a unique way on a given sampling set. Therefore, again the blue-noise sampling patterns that are characterized by small values of $R_{\boldsymbol{s}}$ represent a better option than arbitrary random sampling, which leads to large values of $R_{\boldsymbol{s}}$.

It is important to point out that, under the conditions stated in Theorem
\ref{theorem_Lambdapartitions1}, Theorems~\ref{theorem_lambdaremov_bncostfunc} and~\ref{theorem_Ks}
show that the reduction of the redness is a desirable attribute  for any sampling pattern, which is something that can be considered with other sampling approaches. Additionally, the tightness of the inequalities depends on the graph structure which makes these results stronger in some families of graphs.
\subsection{Stability and blue-noise sampling Sets} 
The selection of a sampling set, $S$, is not only associated to a possible unique representation of a signal but also to the stability of its reconstruction when the samples are corrupted by noise, or when the signal considered is not exactly bandlimited. This
stability can be measured considering the condition of the matrix $\mathbf{U}_{k}(S,:)$, which is the matrix obtained by sampling $\mathbf{U}_{k}$ on the rows indicated by $S$~\cite{ortega_proxies}. Several cost functions can be formulated in terms of the condition
of $\mathbf{U}_{k}(S,:)$, such that their minimum or maximum values are given by the sampling set that provides the best condition~\cite{ortega_proxies}.

The measures of stability provided in~\cite{ortega_proxies} can be equivalently obtained from a formal general definition of stability for sampling sets in arbitrary spaces~\cite{eldarsampling}. In particular, recalling the definition of stability presented
in~\cite{eldarsampling} for general sampling schemes, we can say that $PW_{\omega}(G)$ posses a stable sampling expansion or reconstruction on $S\subset V(G)$ if there exists $C\geq1$
%$C\in\mathbb{R}_{+}$ 
such that $\Vert\boldsymbol{x}(S^{c})\Vert_{2}^{2}\leq (C-1)\Vert\boldsymbol{x}(S)\Vert_{2}^{2}\quad\forall\boldsymbol{x}\in PW_{\omega}(G)$. The value of $C$ provides a measure of stability associated to $S$; the larger the value of $C$ the less
stability we have. 
%This condition could be equivalently stated
%as $\Vert\boldsymbol{x}(S^{c})\Vert_{2}^{2}\leq C^{'}\Vert\boldsymbol{x}(S)\Vert_{2}^{2}\quad\forall\boldsymbol{x}\in PW_{\omega}(G)$, with $C^{'}=C-1$. 
In the following theorem we provide an estimate of $C-1$ in terms of $\Lambda_{S^{c}}$.
\begin{theorem}
	Let $\boldsymbol{x}\in PW_{\omega}(G)$. Then, if $\Lambda_{S^{c}}>\tilde{\omega}$, it follows that
	\begin{equation}
	\Vert\boldsymbol{x}(S^{c})\Vert_{2}^{2}\leq\frac{\left(\frac{\tilde{\omega}^{r}}{\Lambda_{S^{c}}}\right)^{2}}{1-\left(\frac{\tilde{\omega}^{r}}{\Lambda_{S^{c}}}\right)^{2}}\Vert\boldsymbol{x}(S)\Vert_{2}^{2} \quad r\in\mathbb{R}_{+}
	\label{eq_stabilitytheorem}
	\end{equation}
	where $\tilde{\omega}$ is the bandwidth of $\boldsymbol{x}_{1}$, $\boldsymbol{x}_{1}(S^{c})=\boldsymbol{x}(S^{c})$ 
	and $\boldsymbol{x}_{1}(S)=0$.\\
	Proof: See Appendix \ref{appendix_theorem_recstability}
	\label{theorem_recstability}
\end{theorem}
From this theorem, it is important to point out that finding the sampling set $S$ for which $\Lambda_{S^{c}}$ is maximum not only provides a unique representation of a bandlimited signal, but also provides the sampling set in which the 
highest stability is achieved. This result is consistent with the findings in~\cite{ortega_proxies}.

As it was stated in Theorem \ref{theorem_lambdaremov_bncostfunc}, patterns with a low redness promote large values of $\Lambda_{S^{c}}$, therefore blue-noise
sampling patterns not only promote uniqueness of the representation but also stability in the reconstruction. 
 
\subsection{Connection with other works}
The implications and properties of spreading the sampling points as far as possible from each other on non Euclidean domains were formally established by Pesenson in~\cite{pesenson_manifolds1} considering functions on compact Riemmanian manifolds.  In~\cite{bnsurfaces1, bnsurfaces2, bnsurfaces3} the concept of blue-noise was used for the sampling of surfaces embeded in $\mathbb{R}^{3}$
for applications in computer graphics. This last result  can be considered an application of the results in~\cite{pesenson_manifolds1} for  two-dimensional manifolds embedded in $\mathbb{R}^{3}$. It is important to point out  that the results
in~\cite{bnsurfaces1, bnsurfaces2, bnsurfaces3} rely on the mapping that can be established between the surface and a subset of the 2-dimensional Euclidean domain, but they do not offer any insight of how to deal with the problem in higher 
dimensions. In~\cite{krause} a method is proposed for the optimal location of sensors in Euclidean spaces. Exploiting the concepts of entropy, mutual information and Gaussian processes (GPs), the problem of selecting a subset of sensors among a predefined set of
discrete positions, defined 
on a grid of an $n-$dimensional Euclidean space, is addressed. To deal with a large number of possible sensor locations some relaxations based on \textit{lazy evaluations} and local structure 
of (GPs) are used.
%a method is proposed for the optimal location of sensors on subsets of $\mathbb{R}^{2}$ using tools from Gaussian processes, in particular 
%the concept of mutual information is exploited and the solutions obtained can be connected with the ideal sampling points for functions on manifolds stated in~\cite{pesenson_manifolds1} but considering a 2-dimensional manifold with no curvature. 

In a  different context, and before the emergence of graph signal processing, functions defined on the vertices of a graph have been considered under the concept of
\textit{fitness landscapes}~\cite{biyikoglu2007laplacian, stadler1996}, which were introduced  as a tool for the study of molecular evolution. In this context, the \textit{length} of the autocorrelation function has been useful for the analysis of a landscape. In particular, the \textit{correlation length} of a landscape, $\boldsymbol{x}$, on a $K$-regular graph is given by~\cite{stadler1996}
\begin{equation}
\ell_{\boldsymbol{x}}=\frac{K}{\Vert\hat{\boldsymbol{x}} \Vert_{2}^{2}}\sum_{\ell=2}^{N}\frac{\hat{\boldsymbol{x}}(\ell)^{2}}{\mu_{\ell}}.
\label{eq_corrlengthlandscape}
\end{equation}
The values of eqn.~(\ref{eq_corrlengthlandscape}) provide an indication about how correlated are a given set of samples of the landscape obtained using a random walk. As can be observed in eqn.~(\ref{eq_corrlengthlandscape}) this is proportional to the redness of $\boldsymbol{x}$. In this context, it is possible to conceive blue-noise sampling patterns on graphs as landscapes with a low length correlation.

\section{Generating Blue-Noise Sampling Sets}
%%%%%%%%%%%%%%%%%%%%%%%%%%%%%%%%%%%%%%%%%%%%%%%%%%%%%%%%%%%%%%%%%%%%%%%%%%%%%%%%%%%%%%%%%%%%%%%%%%%%%%%%%%%%%%%%%%%%%%%%%%%%%%

\begin{figure*}
	\centering\includegraphics[scale=1,angle=0]{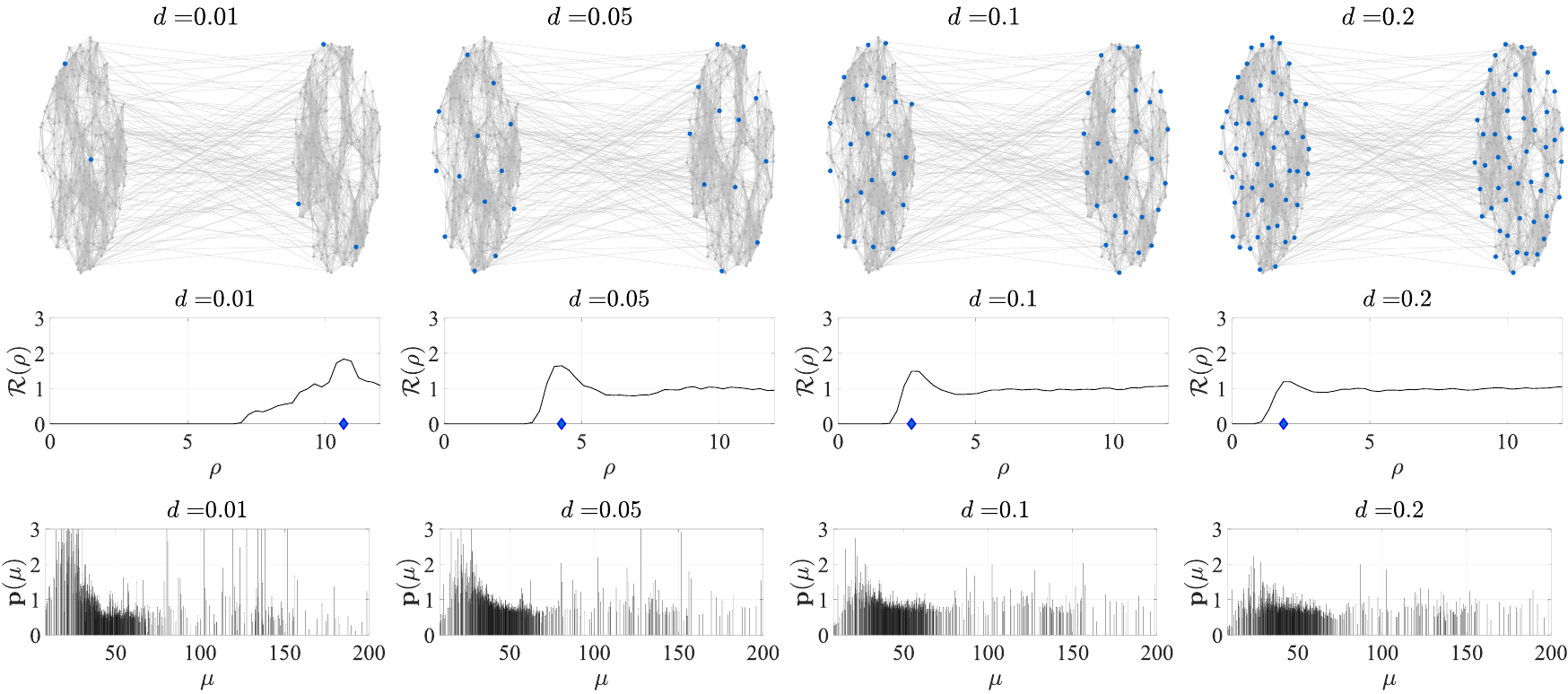} 
	\caption{Void and cluster blue-noise sampling patterns for different intensities $d$ for a community graph. First row: Localization on the graph of the nodes selected in a blue-noise sampling pattern. Second row: The pair correlation function $\mathcal{R}(\rho)$ for the sampling patterns indicating with a diamond marker the value of $\lambda_{b}$. Third row: Power spectral density for the different blue-noise sampling patterns.}
	\label{fig_BNbigpicture2community}
\end{figure*} 
%%%%%%%%%%%%%%%%%%%%%%%%%%%%%%%%%%%%%%%%%%%%%%%%%%%%%%%%%%%%%%%%%%%%%%%%%%%%%%%%%%%%%%%%%%%%%%%%%%%%%%%%%%%%%%%%%%%%%%%%%%%%%%
%%%%%%%%%%%%%%%%%%%%%%%%
\begin{figure}
	\centering\includegraphics[scale=0.5,angle=0]{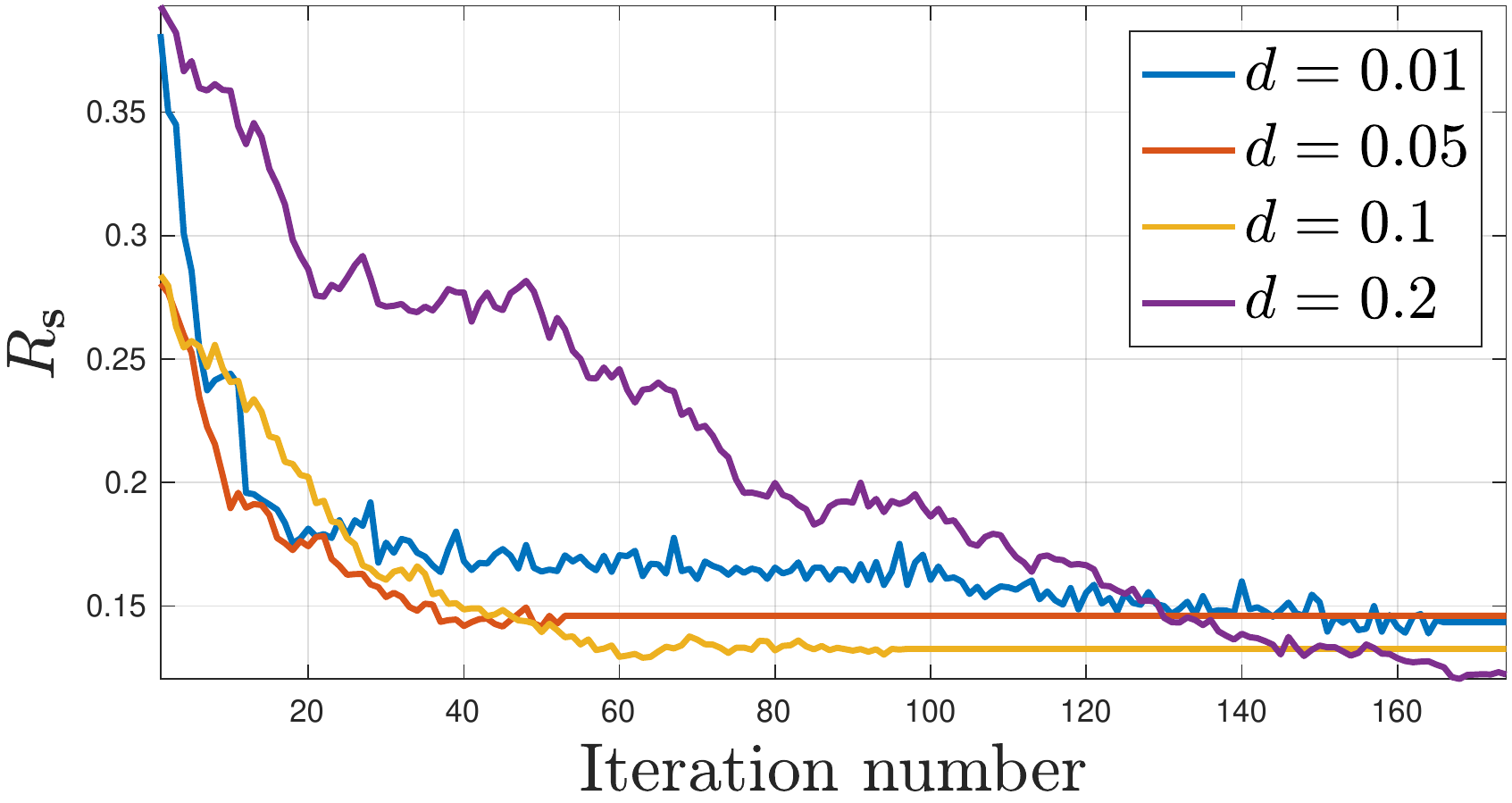} 
	\caption{Illustration of the redness, $R_{\boldsymbol{s}}=\frac{1}{m}\sum_{\ell=2}^{N}\frac{\hat{\boldsymbol{s}}(\ell)^{2}}{\mu_{\ell}}$, of the void and cluster blue-noise sampling patterns on a sensor network with $N=2000$ nodes, considering different densities.}
	\label{fig_VCredness}
\end{figure}
%%%%%%%%%%%%%%%%%%%%%%%%%%%%%%%%%%%%%%%%%%%%%%%%%%%%%%%%%%%%%%%%%%%%%%%%%%%%
Given that blue-noise graph signal sampling promotes the finding of good sampling sets, it is natural to ask how such sampling patterns can be generated. An algorithm that has been particularly successful in digital halftoning and that intuitively translates to graphs is the Void-And-Cluster (VAC) algorithm, introduced by Ulichney~\cite{ulichney_voidcluster}. VAC allows for the construction of artifact-free homogeneous dithering patterns by iteratively measuring the concentration of minority pixels in a binary halftone image, using a gaussian low-pass filter, and swapping the minority pixels in the area of highest concentration with the non-minority pixel in the area of lowest concentration. The adaptation of this algorithm to sampling signals on graphs consists roughly speaking of the sequential computation of distances between sampling points in such a way that  points with short geodesic distances between them are relocated trying to put them far from each other.

In order to exploit the above principle for the selection of sampling nodes on a graph, a Gaussian kernel  $\displaystyle \mathbf{K}(u,v)=exp(-\mathbf{\Gamma}(u,v)^{2}/\sigma)$ is evaluated on the set of geodesic distances, $\mathbf{\Gamma}$. This provides a new set of distances that can be tuned according to the parameter, $\sigma$, where a small value of $\mathbf{\Gamma}(u,v)$ leads to a value of $\mathbf{K}(u,v)$ that is close to unity while a large value of $\mathbf{\Gamma}(u,v)$ leads to a value of $\mathbf{K}(u,v)$ close to zero. As a measure of how homogeneously distributed the sampling points are, the sum of all distances from one node to the others via the kernel $\mathbf{K}$ is calculated as $\mathbf{c}=\mathbf{K}\mathbf{1}_{N\times 1}$. With this, an initial sampling pattern is generated selecting the $m$ components of $\mathbf{c}$ at random, where $m=dN$ is the number of 1's in the sampling pattern with density $d$. 

\begin{algorithm}[t!]
	\caption{Void and cluster algorithm for graphs}
	\begin{algorithmic}[2]
		\renewcommand{\algorithmicrequire}{\textbf{Input:}} 
		\renewcommand{\algorithmicensure}{\textbf{Output:}}
		\REQUIRE $m$: number of samples, $\sigma$, \texttt{NumIter}.
		\ENSURE  $\boldsymbol{s}$: sampling pattern
		\\ \textit{Initialisation} : $\boldsymbol{s}=\mathbf{0}$, \texttt{IndA=-1}, \texttt{IndB=-1}.
		\STATE Calculate $\displaystyle \mathbf{K}(i,j)=e^{-\frac{\mathbf{\Gamma}(i,j)^{2}}{\sigma}}$ for all $1\leq i,j\leq N$.
		\STATE $\mathbf{c}=\mathbf{\mathbf{K}}\mathbf{1}_{N\times 1}$.
		\STATE Get $\mathcal{M}$ as  $m$ nodes selected at random.
		\STATE $\boldsymbol{s}(\mathcal{M})=1$. 
		\FOR {$r=1:1:\texttt{NumIter}$}
		\STATE $\mathbf{c}(\supp(\boldsymbol{s}))=\sum \mathbf{K}(\supp(\boldsymbol{s}),\supp(\boldsymbol{s}))$.
		\STATE $\mathbf{c}(\supp(\boldsymbol{s})^{c})=\sum \mathbf{K}(\supp(\boldsymbol{s}),\supp(\boldsymbol{s})^{c})-\tau$.
		
		\STATE $\boldsymbol{s}\left(\arg\max_{i}\{ \mathbf{c}(i)\} \right)=0$.
		\STATE $\boldsymbol{s}\left(\arg\min_{i}\{ \mathbf{c}(i)\} \right)=1$.
		
		\IF {\texttt{IndA=}$\arg\max_{i}\{ \mathbf{c}(i)\}$ and \texttt{IndB=}$\arg\min_{i}\{ \mathbf{c}(i)\}$}
		\STATE break
		\ELSE
		\STATE \texttt{IndA=}$\arg\min_{i}\{ \mathbf{c}(i)\}$.
		\STATE \texttt{IndB=}$\arg\max_{i}\{ \mathbf{c}(i)\}$.
		
		\ENDIF   
		\ENDFOR
		\RETURN $\boldsymbol{s}$
	\end{algorithmic} 
	\label{alg_voidandcluster}
\end{algorithm}

 The components of $\mathbf{c}$ whose index is given by the location of the 1's in $\boldsymbol{s}$, are then updated to be $\mathbf{c}(\supp(\boldsymbol{s}))=\sum \mathbf{K}(\supp(\boldsymbol{s}),\supp(\boldsymbol{s}))$, where $\sum\mathbf{K}(A,B)$ is defined by
 \begin{equation}
\sum\mathbf{K}(A,B)=\sum_{a_{i},b_{j}}\mathbf{K}(a_{i},b_{j})\quad a_{i}\in A, b_{j}\subset{B}.
 \end{equation}
 The remaining components of $\mathbf{c}$ are updated according to $\mathbf{c}(\supp(\boldsymbol{s})^{c})=\sum \mathbf{K}(\supp(\boldsymbol{s}),\supp(\boldsymbol{s})^{c})-\tau$, where $\tau$ is selected as a large scalar value. With this update the distances
 between sampling points in the pattern are represented as positive quantities without adding the distances to other nodes. The distance between $\supp(\boldsymbol{s})$ and $\supp(\boldsymbol{s})^{c}$ is then represented with a negative value. Now the index of 
 the component of $\mathbf{c}$ with the highest value will indicate the sampling point that is closest to the other sampling points, and then the value of $\boldsymbol{s}$ at that index is forced to be $0$ whereas in the index where $\mathbf{c}$ is minimum,
 $\boldsymbol{s}$ is forced to be 1. Notice that the role of $\tau$ is to make sure that always $\mathbf{c}(\supp(\boldsymbol{s})^{c})<0$ and a variety
 of values for $\tau$ would serve this purpose. Taking into account that $\sum\mathbf{K}(\supp(\boldsymbol{s}),\supp(\boldsymbol{s})^{c})\leq N$,  it is possible to select $\tau$ as any value such that $\tau>N$. 

Repeating the above process iteratively, it is possible to achieve a sampling pattern with no clusters of 1s that exhibits a homogeneous distribution on $V(G)$. The details of the VAC algorithm can be appreciated in Algorithm~\ref{alg_voidandcluster} with
example sampling patterns using VAC depicted in Fig.~\ref{fig_BNbigpicture2sensor}  for the Sensor Network graph and in Fig.~\ref{fig_BNbigpicture2community} for a community graph.  From observation, one can see a clear distinction with respect to random
sampling when it comes to the nodes distribution of the sampling set. The spatial and spectral blue-noise-like behavior is obtained as a byproduct of the algorithm.

At this point, we note that the value of $\sigma$ in the kernel $exp(-\mathbf{\Gamma}(u,v)^{2}/\sigma)$ plays a critical role in VAC as it defines which sampling nodes are close enough to another one in order to produce a relocation of the $1$'s in the
sampling pattern. Taking into account the definition of $\lambda_{b}$ presented in previous sections, it is possible to establish a natural connection between $\sigma$ and $\lambda_{b}$. In order to do so, we note that if the blue-noise sampling pattern
is ideally distributed on the set of nodes, $V(G)$, then when $u$ and $v$ are sampling nodes it follows that $exp(-\mathbf{\Gamma}(u,v)^{2}/\sigma)\approx 0$ if $\mathbf{\Gamma}(u,v)\geq\lambda_{b}$. This criteria is considered to be satisfied
when $\sigma=\lambda_{b}^{2}/ln(10)$, i.e selecting  $\sigma$ in this way the exponential reaches a value of $0.1$ when $\mathbf{\Gamma}(u,v)=\lambda_{b}$. The number of iterations \texttt{NumIter} is selected as a multiple of $N$. In the numerical
experiments performed, we have found that choosing $\texttt{NumIter}=N$ is enough for the algorithm to reach a stationary behavior.
\begin{figure*}
	\centering\includegraphics[scale=1,angle=0]{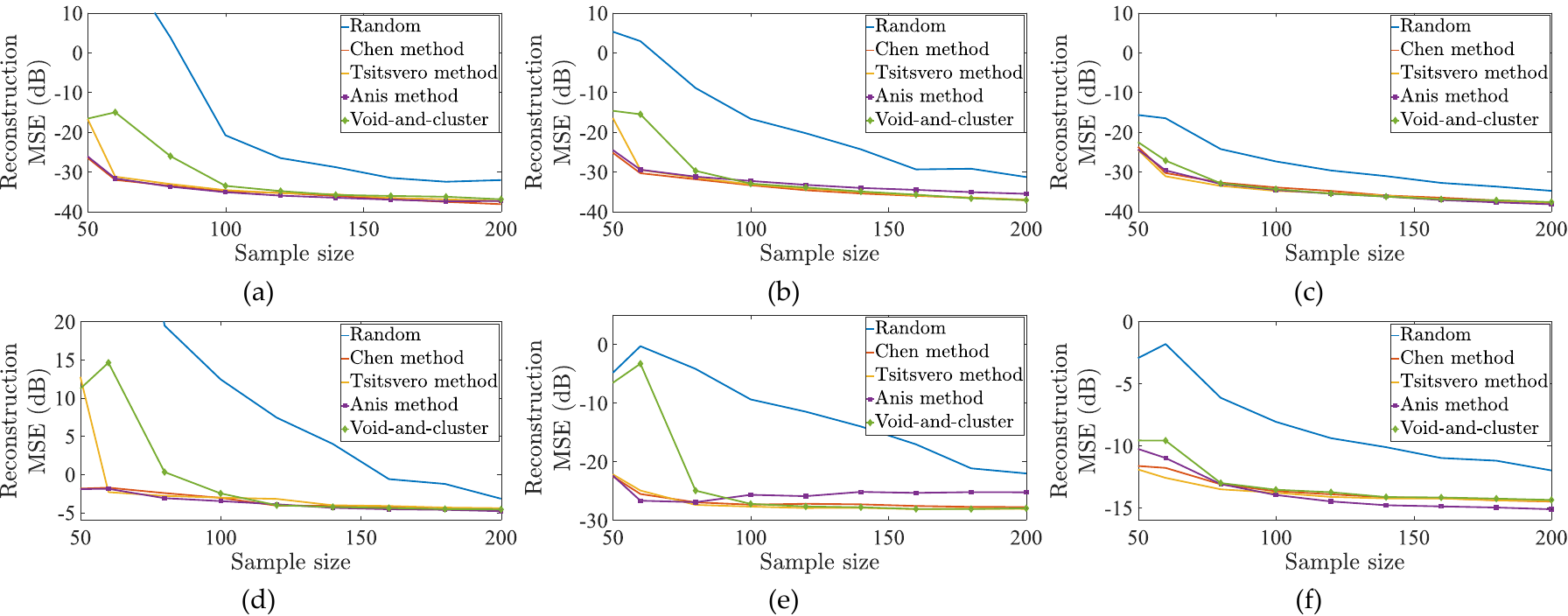} 
	\caption{Averaged MSE using the reconstruction stated in (\ref{eq_basicrec}) vs the sampling rate considering the reconstruction of 100 different signals from its samples using several sampling schemes and considering several graphs: (a) The graph $G_{1}$ and the signal model SM1. (b) The graph $G_{2}$ and the signal model SM1. (c) The graph $G_{3}$ and the signal model SM1. (d) The graph $G_{1}$ and the signal model SM2. (e) The graph $G_{2}$ and the signal model SM2. (f) The graph $G_{3}$ and the signal model SM2.}
	\label{fig_finalresultsall}
\end{figure*}
 \begin{figure*}
 	\centering\includegraphics[scale=1,angle=0]{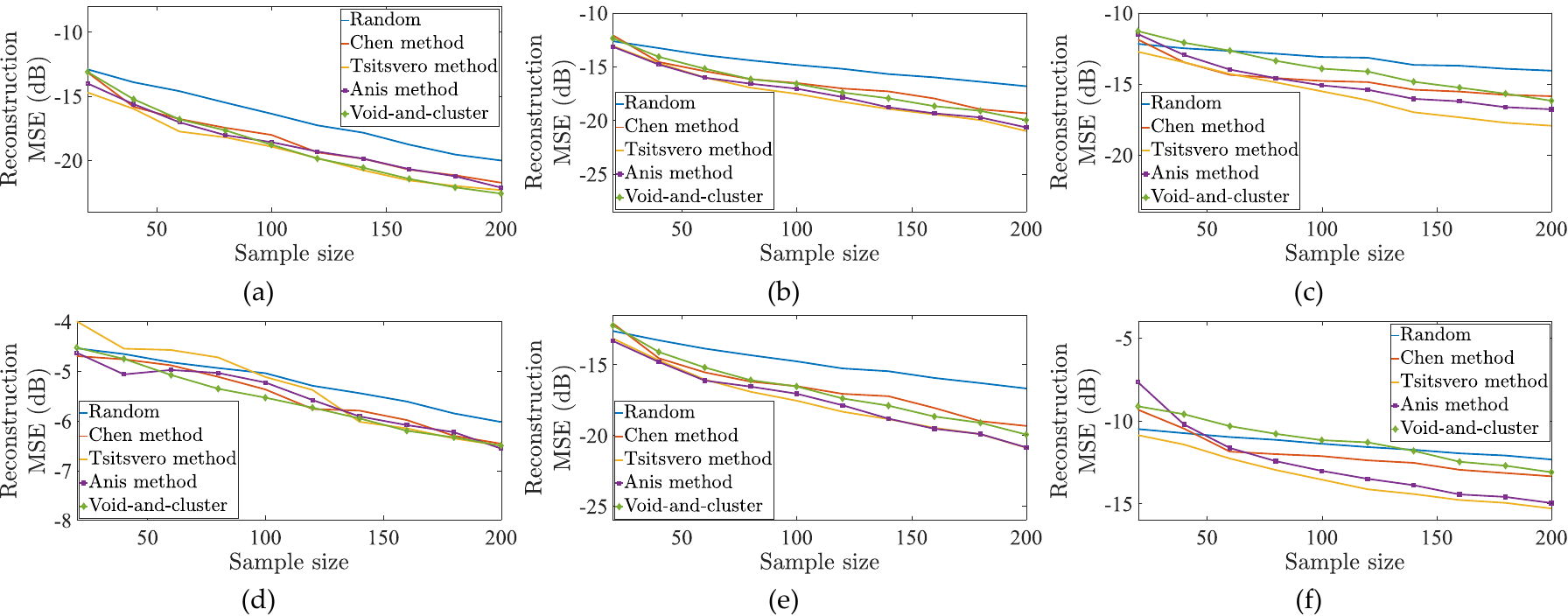} 
 	\caption{Averaged MSE using the reconstruction method proposed in~\cite{pesenson2009} vs the sampling rate considering the reconstruction of 100 different signals from its samples using several sampling schemes and considering several graphs: (a) The graph $G_{1}$ and the signal model SM1. (b) The graph $G_{2}$ and the signal model SM1. (c) The graph $G_{3}$ and the signal model SM1. (d) The graph $G_{1}$ and the signal model SM2. (e) The graph $G_{2}$ and the signal model SM2. (f) The graph $G_{3}$ and the signal model SM2.}
 	\label{fig_finalresultsall2}
 \end{figure*}
%  \begin{figure*}
%  	\centering\includegraphics[scale=1,angle=0]{./figures/results_papers_Puy.pdf} 
%  	\caption{Averaged MSE vs the sampling rate considering the reconstruction of 100 different signals from its samples using several sampling schemes and considering several graphs: (a) The graph $G_{1}$ and the signal model SM1. (b) The graph $G_{2}$ and the signal model SM1. (c) The graph $G_{3}$ and the signal model SM1. (d) The graph $G_{1}$ and the signal model SM2. (e) The graph $G_{2}$ and the signal model SM2. (f) The graph $G_{3}$ and the signal model SM2.}
%  	\label{fig_finalresultsall3}
%  \end{figure*}
 As indicated in Fig.~\ref{fig_VCredness}, there is a clear reduction of the redness of the patterns as they get better distributed on the nodes of the graph. It is important to mention that the number of iterations required for the redness to drop to its minimum value increases as the value of $d$ increases. This is related with the fact that, as $d$ is reduced, there are more possibilities for the relocation of the $1$'s in the sampling pattern.
 
%In traditional halftoning void and cluster provide the best uniformity and spreading of the sampling points, however its computational cost motivated the development of fast algorithms, like error diffusion,  that provide a \textit{good enough} spreading

\section{Experiments}
\begin{figure}
	\centering\includegraphics[scale=1,angle=0]{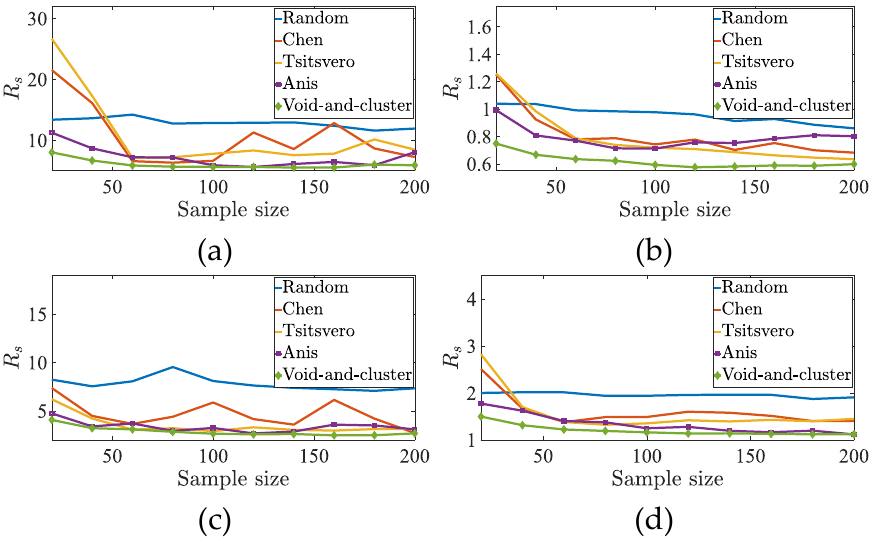} 
	\caption{Illustration of the redness $R_{\boldsymbol{s}}=\frac{1}{m}\sum_{\ell=2}^{N}\frac{\hat{\boldsymbol{s}}(\ell)^{2}}{\mu_{\ell}}$ for the sampling patterns generated by different sampling approaches on different graphs. (a) Swiss roll graph; 
	(b) Sensor network graph; (c) Sphere graph; (d) Bunny graph.}
	\label{fig_rednessallmethods}
\end{figure}

In order to evaluate the benefits of blue-noise sampling, a set of numerical experiments is performed comparing the obtained results against state of the art techniques. The simulations are performed considering different graphs and signal models.  The experiment 
is described by the following steps:
\begin{itemize}[leftmargin=*]
	\item For each graph model, a set of 100 signals is generated according to the  specific signal models selected. 
    \item Each signal is sampled by means of different sampling schemes.
	\item The signal reconstructed from the samples is compared to the original one, and its mean squared error  (MSE) is calculated.
	\item The values of the MSE are averaged over 100.
\end{itemize}
The schemes of sampling considered for the experiment are the following:
\begin{itemize}[leftmargin=*]
	\item Blue noise sampling by void and cluster.
	\item Sampling scheme proposed by Chen et. al.~\cite{chensamplingongraphs}.
	\item Sampling scheme proposed by Anis et. al.~\cite{ortega_proxies}.
	\item Sampling scheme proposed by Tsitsvero et al.~\cite{tsitsverobarbarossa}.
\end{itemize}
The signal models are: 
 \begin{itemize}[leftmargin=*]
 	\item Signal model 1 (SM1):  A random signal of bandwidth $k=50$, where the Fourier coefficients are generated from the Gaussian distribution $\mathcal{N}(1,0.5^{2})$.  The samples captured are contaminated with additive Gaussian noise such that the Signal to Noise Ratio is $SNR=20$dB.
 	\item Signal model 2 (SM2): A random signal with Fourier coefficients generated from the Gaussian distribution $\mathcal{N}(1,0.5^{2})$. This signal is modulated on the spectral axes by $h(\mu)$, where
 	\begin{equation}
 	h(\mu)=\left\lbrace 
 	\begin{array}{ccc}
 	1                                             &    \text{If}   & \mu\leq\mu_{50}\\
 	e^{-4(\mu-\mu_{50})} &    \text{If}   & \mu>\mu_{50}\\
 	\end{array}
 	\right.
 	\end{equation}	
 \end{itemize}
The graphs considered in the simulations are different from each other in their nature and represent typical graphs that can be found in different scenarios and applications. The graph models used are:
\begin{itemize}[leftmargin=*]
	\item Graph $G_{1}$: A random sensor network with $N=1000$ nodes. The weights in the graph are given by the Euclidean distance between points. The maximum number of neighbors for each node is 6.
	\item Graph $G_{2}$: A community graph with $N=1000$ nodes, 16 communities generated using the GSP toolbox~\cite{gsptoolboxref}.
	\item Graph $G_{3}$: A Barab\'asi-Albert random network~\cite{barabasi2016network} with $N=1000$ nodes.
\end{itemize}
The reconstructions are performed by means of eqn. (\ref{eq_basicrec}) and by the interpolation splines proposed in~\cite{pesenson2009} and implemented in~\cite{gsptoolboxref}. In Figs.~\ref{fig_finalresultsall} and~\ref{fig_finalresultsall2}, the performance
of different algorithms can be appreciated including VAC sampling. Notice that the decay rate of the error curves show
consistently the benefits of blue-noise sampling. The results obtained using VAC are close to the ones
obtained in~\cite{ortega_proxies}. Additonally, in Fig.~\ref{fig_rednessallmethods}, the redness of the sampling patterns obtained by different techniques are presented considering different graphs. It is possible to see how low redness is 
a characteristic attribute of good sampling patterns.

\section{Conclusion}
Blue-noise sampling on graphs is defined based on the traditional blue-noise model associated with digital halftones. The properties and benefits of blue-noise sampling on graphs are linked with theoretical results related to
uniqueness sets in sampling, showing why blue-noise patterns promote good sampling sets. We also extended void and cluster, a popular halftoning scheme to generating blue-noise sampling sets on graphs. Numerical tests on different graphs corroborate the good qualities of sampling
with blue-noise. We specified conditions under which the traditional relationship between vertex-domain spreading and frequency behavior is preserved. We further 
note that the results obtained in this work can be extended for specific families of graphs. The delimitation of the properties for the graphs under consideration could lead to sharper bounds and could allow the definition of other quantities
extensively used in halftoning, like the principal frequency.

An overlooked benefit to blue-noise sampling is the wealth of computationally efficient algorithms used in digital halftones that can be extended to graph sampling without spectral estimation, namely error-diffusion where the produced halftone
patterns conform to the local spectral content of the image to optimally preserve salient features like edges, gradients, flood fills, etc. Also, a very valuable attribute of error-diffusion that is not widely recognized outside the halftoning community 
is that error-diffusion can be trained to produce arbitrary spectral profiles (blue-noise, green-noise, etc) and even designed to match the dither patterns produced by other means, including ones with high computational
complexity~\cite{bluenoiseLaumagazine,lauarce_generrmultch}.  For graph signal sampling, this opens up the possibility of error-diffusion algorithms trained to mimic sampling algorithms based on spectral estimation and vertex-domain characteristics. We consider 
that an interesting topic for future research would be the analysis and implications of blue-noise sampling on graphs for signals that are bandlimited but not necessarily low pass~\cite{Lau99,735449}. This could provide a generalization of the results that were stated 
in this work. 

It is important to point out that blue-noise sampling promotes large values of $\Lambda_{S^{c}}$, but there is not a guarantee about reaching the maximum value of $\Lambda_{S^{c}}$. For this reason the stability is affected when the value of $\Lambda_{S^{c}}$ is
not large enough, which also happens when $m$ is not large enough. This aspect is something that can be improved in future works adding additional constraints to the method used to generate blue-noise sampling patterns. 
\appendices

\section{Proof of Theorem~\ref{theorem_Lambdapartitions1} }
\label{appendix_prooftheorem_Lambdapartitions1}
\begin{proof}
	As stated in~\cite{chung1997spectral} (page 168), by means of Sobolev inequalities, it is possible to state that
	\begin{equation}
	\mu_{1,j}\geq C_{\delta_{j}}\frac{1}{vol(\Omega_{j})^{\frac{2}{\delta_{j}}}}   
	\label{eq_tempdirichletsobolev}
	\end{equation}
	where $\delta_{j}$ is the isoperimetric dimension of $\Omega_{j}$ and $C_{\delta_{j}}$ is a constant that depends only on $\delta_{j}$. Taking into account the definition of $\Lambda_{\mathcal{P}}$  in Theorem~\ref{pesentheorem2019_3}
	we do have that
	\begin{equation*}
	\Lambda_{\mathcal{P}}>\min\left\lbrace  
	 \frac{C_{\delta_{1}}}{vol(\Omega_{1})^{\frac{2}{\delta_{1}}}}    ,\ldots ,  \frac{C_{\delta_{\vert \mathcal{P}\vert}}}{vol(\Omega_{\vert \mathcal{P}\vert})^{\frac{2}{\delta_{\vert \mathcal{P}\vert}}}}
	\right\rbrace
	\end{equation*}
	if $\delta_{1}=\delta_{2}=\ldots =\delta_{\vert\mathcal{P}\vert}=\delta$, then it follows that
	\begin{equation*}
	\Lambda_{\mathcal{P}}>\min\left\lbrace  
    \frac{C_{\delta}}{vol(\Omega_{1})^{\frac{2}{\delta}}} ,\ldots , \frac{C_{\delta}}{vol(\Omega_{\vert \mathcal{P}\vert})^{\frac{2}{\delta}}}
	\right\rbrace
	\end{equation*}
\end{proof}

\section{Proof of Theorem~\ref{theorem_Lambdapartitions2} }
\label{appendix_prooftheorem_Lambdapartitions2}
\begin{proof}
	In order to simplify notation lets represent $x_{i}=vol(\Omega_{i})^{2/\delta}/C_{\delta}$ and lets consider the optimization problem
	\begin{equation}
	\begin{aligned}
	& \underset{\{x_{1},x_{2}, \ldots, x_{\vert\mathcal{P}\vert}\}}{\text{maximize}}
	& & \min \{ 1/x_{1}, 1/x_{2},\ldots, 1/x_{\vert\mathcal{P}\vert} \} \\
	& \text{subject to}
	& & \sum_{i=1}^{\vert\mathcal{P}\vert}x_{i}=c_{1},\quad  x_{i}>0~\forall i \\
	\end{aligned}
	\label{eq_optzproof}
	\end{equation}
	where $c_{1}$ is a constant. Now, taking into account that
	\begin{equation*}
	\min \{ 1/x_{1}, 1/x_{2},\ldots, 1/x_{\vert\mathcal{P}\vert} \}\sum_{i=1}^{\vert \mathcal{P}\vert}x_{i}\leq \sum_{i=1}^{\vert \mathcal{P}\vert}x_{i}\frac{1}{x_{i}}=\vert\mathcal{P}\vert
	\end{equation*}
	\begin{equation*}
	\min \{ 1/x_{1}, 1/x_{2},\ldots, 1/x_{\vert\mathcal{P}\vert} \}\leq  \vert\mathcal{P}\vert/c_{1}.
	\end{equation*}
	Then, the maximum value of the objective function in eqn.~(\ref{eq_optzproof}) is $\vert\mathcal{P}\vert/c_{1}$. Let $(x_{1}^{*},x_{2}^{*},\ldots, x_{\vert\mathcal{P}\vert}^{*})$ the optimal solution of~(\ref{eq_optzproof}), then it follows that
	$\vert\mathcal{P}\vert/c_{1}\leq 1/x_{i}^{*}$. 
	Lets assume there exists a subset of indexes $\{ j_{1}, j_{2},\ldots, j_{q}\}\subset \{1,\ldots, \vert\mathcal{P}\vert \}$ such that $\vert\mathcal{P}\vert/c_{1}< 1/x_{j_{r}}^{*}$ which implies $x_{j_{r}}^{*}<c_{1}/\vert\mathcal{P}\vert$. Then it follows that
	\begin{equation}
	\sum_{i=1}^{\vert \mathcal{P}\vert}x_{i}=\left( \vert\mathcal{P}\vert - q\right)\frac{c_{1}}{\vert\mathcal{P}\vert}+\sum_{j_{r}}x_{j_{r}}<c_{1}
	\end{equation}
	which is a contradiction. Therefore $x_{i}^{*}=c_{1}/\vert\mathcal{P}\vert$ which implies $x_{1}^{*}=x_{2}^{*}=\ldots=x_{\vert\mathcal{P}\vert}^{*}$.
\end{proof}

\section{Proof of Theorem~\ref{theorem_parsandsamppattern}  }
\label{appendix_theorem_parsandsamppattern}
\begin{proof}
Let $\boldsymbol{s}\in \{ 0,1\}^{N}$ selected according to (\ref{eq_homogcond1}) and (\ref{eq_homogcond2}), then it follows that $\boldsymbol{s}^{\mathsf{T}}\mathbf{L}\boldsymbol{s}=\sum_{j=1}^{\vert\mathcal{P}\vert}\boldsymbol{s}(V(\Omega_{j}))^{\mathsf{T}}\mathbf{L}_{\Omega_{j}}\boldsymbol{s}(V(\Omega_{j})).$
%\begin{equation}
%\boldsymbol{s}^{\mathsf{T}}\mathbf{L}\boldsymbol{s}=\sum_{j=1}^{\vert\mathcal{P}\vert}\boldsymbol{s}(V(\Omega_{j}))^{\mathsf{T}}\mathbf{L}_{\Omega_{j}}\boldsymbol{s}(V(\Omega_{j})).
%\end{equation}
Additionally, directly from the definition of $\mu_{1,j}$ and using the Raleyigth coefficient we have
$\mu_{1,j}\leq \boldsymbol{s}(V(\Omega_{j}))^{\mathsf{T}}\mathbf{L}_{\Omega_{j}}\boldsymbol{s}(V(\Omega_{j}))$.
%\begin{equation}
%\mu_{1,j}\leq \boldsymbol{s}(V(\Omega_{j}))^{\mathsf{T}}\mathbf{L}_{\Omega_{j}}\boldsymbol{s}(V(\Omega_{j}));
%\end{equation}
Therefore
\begin{equation}
\sum_{j=1}^{\vert\mathcal{P}\vert}\mu_{1,j}\leq \sum_{j=1}^{\vert\mathcal{P}\vert}\boldsymbol{s}(V(\Omega_{j}))^{\mathsf{T}}\mathbf{L}_{\Omega_{j}}\boldsymbol{s}(V(\Omega_{j}))=\boldsymbol{s}^{\mathsf{T}}\mathbf{L}\boldsymbol{s}.
\end{equation}
Now, taking into account eqn.~(\ref{eq_tempdirichletsobolev}) we have
\begin{equation}
\vert\mathcal{P}\vert\min_{j}\frac{C_{\delta_{j}}}{vol(\Omega_{j})^{\frac{2}{\delta_{j}}}}\leq \sum_{j=1}^{\vert\mathcal{P}\vert}\mu_{1,j}\leq \boldsymbol{s}^{\mathsf{T}}\mathbf{L}\boldsymbol{s}=
\sum_{\ell=2}^{N}\mu_{\ell}\hat{\boldsymbol{s}}(\ell)^{2}
\end{equation}
and using lemma \ref{lemma_RQniceineq}, we obtain
\begin{equation*}
R_{\boldsymbol{s}}\leq\frac{(\mu_{2}+\mu_{N})^{2}(1-\vert\mathcal{P}\vert/N)^{2}
}{4\mu_{2}\mu_{N}\min\limits_{j}\left\lbrace \frac{C_{\delta_{j}}}{vol(\Omega_{j})^{\frac{2}{\delta_{j}}}}\right\rbrace}.
\end{equation*}
Now, when $\delta=\delta_{1}=\ldots=\delta_{\vert\mathcal{P}\vert}$, it follows that 
\begin{equation}
\frac{C_{\delta}\vert\mathcal{P}\vert}{ vol(\Omega)^{\frac{2}{\delta}}}\leq \sum_{j=1}^{\vert\mathcal{P}\vert}\mu_{1,j}\leq \boldsymbol{s}^{\mathsf{T}}\mathbf{L}\boldsymbol{s}=
\sum_{\ell=2}^{N}\mu_{\ell}\hat{\boldsymbol{s}}(\ell)^{2}
\end{equation}
and from lemma \ref{lemma_RQniceineq}, it follows that 
\begin{equation*}
R_{\boldsymbol{s}}\leq\frac{(\mu_{2}+\mu_{N})^{2}(1-\vert\mathcal{P}\vert/N)^{2}vol(\Omega)^{\frac{2}{\delta}}}{4C_{\delta}\mu_{2}\mu_{N}}.
\end{equation*}
\end{proof}

\section{Redness inequality }
In this section an important and useful lemma used in several proofs is stated.
\begin{lemma}
For any sampling pattern $\boldsymbol{s}: V(G)\rightarrow\{0,1\}$, it follows that	
\begin{equation}
\frac{m\left(1-\frac{m}{N}\right)^{2}}{ \sum_{\ell=2}\mu_{\ell}\hat{\boldsymbol{s}}(\ell)^{2} }\leq R_{\boldsymbol{s}} 
\leq \frac{m(\mu_{2}+\mu_{N})^{2}\left(1-\frac{m}{N}\right)^{2}}{ 4\mu_{2}\mu_{N}\sum_{\ell=2}\mu_{\ell}\hat{\boldsymbol{s}}(\ell)^{2}}
\end{equation}
\label{lemma_RQniceineq}
\end{lemma}
\begin{proof}
By Cauchy inequality we know that 
\begin{multline}
m^{2}\left( 1-\frac{m}{N}\right)^{2}=\left( \sum_{\ell=2}^{N}\sqrt{\mu_{\ell}}\hat{\boldsymbol{s}}(\ell)\frac{1}{\sqrt{\mu_{\ell}}}\hat{\boldsymbol{s}}(\ell)\right)^{2} 
 \leq \\
 \left( \sum_{\ell=2}^{N}\mu_{\ell}\hat{\boldsymbol{s}}(\ell)^{2} \right)
 \left( \sum_{\ell=2}^{N}\frac{1}{\mu_{\ell}}\hat{\boldsymbol{s}}(\ell)^{2}\right)
 \label{eq_tempproofredness1}
\end{multline}
Now, as indicated in~\cite{cauchymasterclass} when $\mu_{\ell}\hat{\boldsymbol{s}}(\ell)>0$ for all $\ell$ we have that
\begin{multline*}
\left( \sum_{\ell=2}^{N}\mu_{\ell}\hat{\boldsymbol{s}}(\ell)^{2} \right)
\left( \sum_{\ell=2}^{N}\frac{1}{\mu_{\ell}}\hat{\boldsymbol{s}}(\ell)^{2}\right) 
\leq \\
\left(\frac{\alpha + \beta}{2\sqrt{\alpha\beta}}\right)^{2}\left(
\sum_{\ell=2}^{N}\sqrt{\mu_{\ell}}\hat{\boldsymbol{s}}(\ell)\frac{1}{\sqrt{\mu_{\ell}}}\hat{\boldsymbol{s}}(\ell) \right)^{2}\\
=\left(\frac{\alpha + \beta}{2\sqrt{\alpha\beta}}\right)^{2}m^{2}\left(1-\frac{m}{N}\right)^{2}
\end{multline*}
with $0<\alpha\leq\mu_{\ell}\leq\beta$. Then, with $\alpha=\mu_{2}$ and $\beta=\mu_{N}$ it follows that 	
\begin{multline}
\frac{4\mu_{2}\mu_{N}}{(\mu_{2}+\mu_{N})^{2}}\left( \sum_{\ell=2}^{N}\mu_{\ell}\hat{\boldsymbol{s}}(\ell)^{2} \right)
\left( \sum_{\ell=2}^{N}\frac{1}{\mu_{\ell}}\hat{\boldsymbol{s}}(\ell)^{2}\right) 
\leq \\
m^{2}\left(1-\frac{m}{N}\right)^{2} 
\label{eq_tempproofredness2}
\end{multline}
combining eqn.~(\ref{eq_tempproofredness1}) and eqn.~(\ref{eq_tempproofredness2}) we obtain
\begin{multline*}
\frac{m\left(1-\frac{m}{N}\right)^{2}}{ \sum_{\ell=2}\mu_{\ell}\hat{\boldsymbol{s}}(\ell)^{2} }\leq  \sum_{\ell=2}^{N}\frac{\hat{\boldsymbol{s}}(\ell)^{2}}{m\mu_{\ell}} 
\leq \frac{(\mu_{2}+\mu_{N})^{2}\left(1-\frac{m}{N}\right)^{2} m}{ 4\mu_{2}\mu_{N}\sum_{\ell=2}\mu_{\ell}\hat{\boldsymbol{s}}(\ell)^{2}}
\end{multline*}
\end{proof}

\section{Proof of Corollary \ref{theorem_partitionfrompattern}}
\label{appendix_theorem_partitionfrompattern}
\begin{proof}
The first part of the proof follows directly from the Definition 1. Now, lets assume  $\Lambda_{\mathcal{P}}>(1+1/\alpha)\omega$  with $\alpha>0$, then according to Theorem~\ref{pesentheorem2019_3} any signal $\boldsymbol{x}\in PW_{\omega}(G)$ is uniquely determined by the values
$\boldsymbol{x}^{\mathsf{T}}\boldsymbol{\xi}_{j}=\sqrt{\vert V(\Omega_{j})\vert}\boldsymbol{x}^{\mathsf{T}}(\boldsymbol{\xi}_{j}\circ\boldsymbol{\xi}_{j})$, therefore if $\boldsymbol{x}(s_{j})=\boldsymbol{x}^{\mathsf{T}}(\boldsymbol{\xi}_{j}\circ\boldsymbol{\xi}_{j})$, $\boldsymbol{x}$ is uniquely determined from $\boldsymbol{x}(s_{i})$.
\end{proof}

\section{Proof of Theorem~\ref{theorem_lambdaremov_bncostfunc}}
\label{appendix_theorem_lambdaremov_bncostfunc}
In order to prove Theorem~\ref{theorem_lambdaremov_bncostfunc}, some preliminary lemmas and theorems are discussed.
\begin{lemma}
For any subset of nodes $S\subset V(G)$ and sampling pattern $\boldsymbol{s}\in\{ 0,1\}^{N}$ with $\supp(\boldsymbol{s})=S$, it follows that
\begin{equation}
vol(S)\geq \frac{m^{2}\left(1-\frac{m}{N}\right)^{2}}{\sum\limits_{\ell=2}^{N}\frac{1}{\mu_{\ell}}\hat{\boldsymbol{s}}(\ell)^{2}}
\label{eq_lemma_volubn}
\end{equation}
where $m=\Vert \boldsymbol{s}\Vert_{0}=\vert S\vert$.
\label{lemma_volubn}
\end{lemma}
\begin{proof}
 Lets consider the Laplacian matrix $\mathbf{L}$. Multiplying  on the left by $\boldsymbol{s}^{\mathsf{T}}$ and on the right hand side by $\boldsymbol{s}$ it follows that $\boldsymbol{s}^{\mathsf{T}}\mathbf{L}\boldsymbol{s}=\boldsymbol{s}^{\mathsf{T}}\mathbf{D}\boldsymbol{s}-\boldsymbol{s}^{\mathsf{T}}\mathbf{W}\boldsymbol{s}$,
% \begin{equation*}
%\boldsymbol{s}^{\mathsf{T}}\mathbf{L}\boldsymbol{s}=\boldsymbol{s}^{\mathsf{T}}\mathbf{D}\boldsymbol{s}-\boldsymbol{s}^{\mathsf{T}}\mathbf{W}\boldsymbol{s}
 %\end{equation*}  	
which  leads to $\sum_{\ell=2}^{N}\mu_{\ell}\hat{\boldsymbol{s}}(\ell)^{2}=vol(S)-\boldsymbol{s}^{\mathsf{T}}\mathbf{W}\boldsymbol{s}$
% \begin{equation*}
%\sum_{\ell=2}^{N}\mu_{\ell}\hat{\boldsymbol{s}}(\ell)^{2}=vol(S)-\boldsymbol{s}^{\mathsf{T}}\mathbf{W}\boldsymbol{s}
% \end{equation*}
and therefore $ \sum_{\ell=2}^{N}\mu_{\ell}\hat{\boldsymbol{s}}(\ell)^{2}\leq vol(S).$
% \begin{equation*}
% \sum_{\ell=2}^{N}\mu_{\ell}\hat{\boldsymbol{s}}(\ell)^{2}\leq vol(S).
% \end{equation*}
 Now, taking into account  the Lemma \ref{lemma_RQniceineq} eqn.~\ref{eq_lemma_volubn} is obtained.
% \begin{equation*}
% vol(S)\geq \frac{m^{2}\left(1-\frac{m}{N}\right)^{2}}{\sum\limits_{\ell=2}^{N}\frac{1}{\mu_{\ell}}\hat{\boldsymbol{s}}(\ell)^{2}}
% \end{equation*}
\end{proof}

\subsection{Proof of Theorem~\ref{theorem_lambdaremov_bncostfunc}}
\begin{proof}
Fuhr and Pesenson~\cite{fuhrpesenson} show that if a subset of nodes $S\subset V(G)$ is removable with constant $\Lambda_{S}$, it follows that $\Lambda_{S}\geq \mu_{D}(S)$, where
$\mu_{D}(S)$ is the 
Dirichlet eigenvalue of the induced subgraph\footnote{Definitions and inequalities about induced subgraphs can be found in~\cite{chung1997spectral}} of $S$. This inequality is tight always that $S\cup bS=V(G)$, where 
$bS$ is the vertex boundary of $S$.

As stated in~\cite{fuhrpesenson}, $\mu_{D}(S)$ satisfy the following inequality
\begin{equation}
\mu_{D}(S)>C_{\delta}\left(\frac{1}{vol(S)}\right)^{2/\delta} 
\end{equation}
where $\delta$ is the isoperimetric dimension of the graph, $C_{\delta}$ is a constant that depends only on $\delta$ and $vol(S)=\sum_{v\in S}D(v,v)$. 

Now, taking into account that $vol(G)=vol(S)+vol(S^{c})$ for 
any $S\subset V(G)$, and the lemma \ref{lemma_volubn}, we have that

\begin{equation}
vol(G)-vol(S)\leq 
vol(G)-\frac{m^{2}\left( 1-\frac{m}{N}\right)^{2}}{\sum\limits_{\ell=2}^{N}\frac{1}{\mu_{\ell}}\hat{\boldsymbol{s}}(\ell)^{2}} 
\end{equation}
%\begin{multline}
%vol(G)-vol(S^{c})\leq 
%vol(G)-\frac{m^{2}\left( 1-\frac{m}{N}\right)^{2}}{\sum\limits_{\ell=2}^{N}\frac{1}{\mu_{\ell}}\hat{\boldsymbol{s}}(\ell)^{2}} 
%\end{multline}
and then
\begin{multline}
C_{\delta}\left( \frac{1}{vol(G)-vol(S)}\right)^{\frac{2}{\delta}}\geq \\
C_{\delta}\left( \frac{\sum_{\ell=2}\frac{1}{\mu_{\ell}}\hat{\boldsymbol{s}}(\ell)^{2}}{vol(G)\sum_{\ell=2}\frac{1}{\mu_{\ell}}\hat{\boldsymbol{s}}(\ell)^{2}-
	m^{2}\left( 1-\frac{m}{N}\right)^{2}}\right)^{\frac{2}{\delta}} .
\end{multline}
Now, taking into account that $\Lambda_{S^{c}}\geq \mu_{D}(S^{c})$, it follows that
\begin{equation*}
\Lambda_{S^{c}}\geq 
C_{\delta}\left( \frac{\sum_{\ell=2}\frac{1}{\mu_{\ell}}\hat{\boldsymbol{s}}(\ell)^{2}}{vol(G)\sum_{\ell=2}\frac{1}{\mu_{\ell}}\hat{\boldsymbol{s}}(\ell)^{2}-
	m^{2}\left( 1-\frac{m}{N}\right)^{2}}\right)^{\frac{2}{\delta}} 
\end{equation*}
\end{proof}

\section{Proof of Theorem~\ref{theorem_Ks}}
\label{appendix_Kstheorem}
In this section the proof of Theorem~\ref{theorem_Ks} is  provided. Before this proof is presented an important lemma is introduced.
\begin{lemma}
Let $\boldsymbol{s}:V(G)\longmapsto\{ 0,1\}^{N}$ a binary signal defined on $V(G)$ and let
$\bar{\boldsymbol{s}}=\mathbf{1}-\boldsymbol{s}$, then it follows that
\begin{equation}
\sum_{\ell=2}^{N}\mu_{\ell}\hat{\bar{\boldsymbol{s}}}(\ell)^{2}=\sum_{\ell=2}^{N}\mu_{\ell}\hat{\boldsymbol{s}}(\ell)^{2}
\end{equation}
\label{lemma_bincompsampattrnFourier}
\end{lemma}
\begin{proof}
Lets consider the Laplacian matrix $\mathbf{L}$ and multiply on the left by $\boldsymbol{s}^{\mathsf{T}}$ and on the right by $\boldsymbol{s}$, it follows that
\begin{equation}
\boldsymbol{s}^{\mathsf{T}}\mathbf{L}\boldsymbol{s}=\left( \mathbf{1}-\bar{\boldsymbol{s}}\right)^{\mathsf{T}}\mathbf{L}\left( \mathbf{1}-\bar{\boldsymbol{s}}\right)=\bar{\boldsymbol{s}}^{\mathsf{T}}\mathbf{L}\bar{\boldsymbol{s}}.
\label{eq_quad_sandsbar}
\end{equation}	
Now, taking into account that 
$\boldsymbol{x}^{\mathsf{T}}\mathbf{L}\boldsymbol{x}= \sum_{\ell=1}^{N}\mu_{\ell}\hat{\boldsymbol{x}}(\ell)^{2}$,
%\begin{multline}
%\boldsymbol{x}^{\mathsf{T}}\mathbf{L}\boldsymbol{x}=\boldsymbol{x}^{\mathsf{T}}\mathbf{U}\mathbf{\Lambda}\mathbf{U}^{\mathsf{T}}
%\boldsymbol{x}=\left( \mathbf{U}^{\mathsf{T}}\boldsymbol{x}\right)^{\mathsf{T}}\mathbf{\Lambda}\left( \mathbf{U}^{\mathsf{T}}\boldsymbol{x}\right)\\
%=\sum_{\ell=1}^{N}\mu_{\ell}\hat{\boldsymbol{x}}(\ell)^{2}
%\end{multline}
it follows that
\begin{equation}
\sum_{\ell=2}^{N}\mu_{\ell}\hat{\bar{\boldsymbol{s}}}(\ell)^{2}=\sum_{\ell=2}^{N}\mu_{\ell}\hat{\boldsymbol{s}}(\ell)^{2}.
\end{equation}
Notice that $\mu_{1}=0$ and consequently the sum can be computed for $\ell\geq 2$.
\end{proof}
\subsection{Proof of Theorem~\ref{theorem_Ks}}
\begin{proof}
Taking into account that 
\begin{equation}
\left( \mathbf{L}\boldsymbol{x}\right)(v)=\sum_{u\in V(G)}\left( \boldsymbol{x}(v)-\boldsymbol{x}(u)\right)\mathbf{W}(v,u)
\end{equation}
 and $w_{S}(v)=\sum_{u\in S}\mathbf{W}(u,v)$. It is possible to infer that
 \begin{equation}
 \left(\mathbf{L}\bar{\boldsymbol{s}}\right)(v)=\left\lbrace 
 \begin{array}{ccc}
w_{S}(v) & \text{if} & v\in S^{c}\\
-w_{S^{c}}(v) & \text{if} & v\in S\\
 \end{array}
 \right.
 \label{eq_prooftheoremks_1}
 \end{equation}	
where $\bar{\boldsymbol{s}}=\mathbf{1}-\boldsymbol{s}$. Now, taking into account eqn.~(\ref{eq_prooftheoremks_1}) and Lemma~\ref{lemma_bincompsampattrnFourier} it follows that
\begin{equation*}
\bar{\boldsymbol{s}}^{\mathsf{T}}\mathbf{L}\bar{\boldsymbol{s}}=\sum_{\ell=2}^{N}\mu_{\ell}\hat{\bar{\boldsymbol{s}}}(\ell)^{2}=\sum_{v\in S^{c}}w_{S(v)}^{2}
\end{equation*}
\begin{equation*}
\sum_{\ell=2}^{N}\mu_{\ell}\hat{\boldsymbol{s}}(\ell)^{2}=K_{S}^{2}+\sum_{v\in\{S^{c}\setminus v^{'}\}}w_{S}(v)^{2}
\end{equation*}
\begin{equation*}
K_{S}=\left( \sum_{\ell=2}^{N}\mu_{\ell}\hat{\boldsymbol{s}}(\ell)^{2}-\sum_{v\in\{S^{c}\setminus v^{'}\}}w_{S}(v)^{2}\right)^{\frac{1}{2}}
\end{equation*}
which leads to
\begin{equation*}
K_{S}\geq\left( \sum_{\ell=2}^{N}\mu_{\ell}\hat{\boldsymbol{s}}(\ell)^{2}-\gamma\right)^{\frac{1}{2}}
\end{equation*}
where $\gamma$ is given by $\gamma=\max_{S,v,v^{'}}\sum_{v\in\{S^{c}\setminus v^{'}\}}w_{S}(v)^{2} $
%\begin{equation}
%\gamma=\max_{S,v,v^{'}}\sum_{v\in\{S^{c}\setminus v^{'}\}}w_{S}(v)^{2}
%\end{equation}
and taking into account Lemma \ref{lemma_RQniceineq}, it follows that
\begin{equation*}
K_{S}\geq\left(
 \frac{m^{2}\left(1-\frac{m}{N}\right)^{2}}{\sum_{\ell=2}^{N}\frac{1}{\mu_{\ell}}\hat{\boldsymbol{s}}(\ell)^{2}}
 -\gamma
 \right)^{\frac{1}{2}}
\end{equation*}
\end{proof}

\section{Proof of Theorem~\ref{theorem_recstability}}
\label{appendix_theorem_recstability}
\begin{proof}
Lets consider $\boldsymbol{x}\in PW_{\omega}(G)$ written as $\boldsymbol{x}=\boldsymbol{x}_{1}+\boldsymbol{x}_{2}$ with $\boldsymbol{x}_{1}(S^{c})=\boldsymbol{x}(S^{c})$, $\boldsymbol{x}_{1}(S)=0$, $\boldsymbol{x}_{2}(S)=\boldsymbol{x}(S)$ and
$\boldsymbol{x}_{2}(S^{c})=0$. From  eqn.~(\ref{eq_Lambdaremsetdef}) we have that 
$\Vert\boldsymbol{x}_{1}\Vert_{2}^{2}\leq (1/\Lambda_{S^{c}}^{2})\Vert\mathbf{L}\boldsymbol{x}_{1}\Vert^{2}$. Now, using Bernstein's inequality~\cite{pesensonams1,fuhrpesenson} we get 
$\Vert\boldsymbol{x}_{1}\Vert_{2}^{2}\leq(1/\Lambda_{S^{c}}^{2})\tilde{\omega}^{2r}\Vert\boldsymbol{x}_{1} \Vert_{2}^{2}$. From this we have $\Vert\boldsymbol{x}_{1}\Vert_{2}^{2}\leq(1/\Lambda_{S^{c}}^{2})\tilde{\omega}^{2r}\left(\Vert\boldsymbol{x}_{1}\Vert_{2}^{2}+\Vert\boldsymbol{x}_{2} \Vert_{2}^{2}\right)$ leading to eqn.~(\ref{eq_stabilitytheorem}).
\end{proof}

\bibliography{bibliography}
\bibliographystyle{unsrt}

\ifCLASSOPTIONcaptionsoff
  \newpage
\fi

% \end{IEEEbiography}
% % insert where needed to balance the two columns on the last page with
% %\newpage

\end{document}